\RequirePackage{amsmath}
\documentclass[epj]{svjour}
\usepackage{graphicx}
\usepackage{comment}
\usepackage{hyperref}
\usepackage{amsmath}
\usepackage{amssymb}
\usepackage{dcolumn}
\usepackage{cuted}
\usepackage{balance}
\usepackage[utf8]{inputenc}

\usepackage{amsmath,amssymb,slashed,dcolumn,amsfonts,slashed,tabulary}
\usepackage{booktabs}
\usepackage{comment}
\usepackage{multirow}
\usepackage{mathptmx}
\usepackage{marginnote}

\newcommand{\btau}{\vec \tau }
\newcommand{\vecq}{\mathbf{q}}
\newcommand{\vecP}{\mathbf{P}}
\newcommand{\vecp}{\mathbf{p}}
\newcommand{\vecr}{\mathbf{r}}

\begin{document}
\title{Uncertainty quantification and falsification of Chiral Nuclear
  Potentials}
% \subtitle{Do you have a subtitle?\\ If so, write it here}
\author{R. Navarro P\'erez \inst{1} \and 
E. Ruiz Arriola\inst{2} }
% \thanks{\emph{Present address:} Insert the address here if needed}%
% }                     % Do not remove
%
% \offprints{}          % Insert a name or remove this line
%
\institute{Department of Physics. San Diego State University. 5500
  Campanile Drive, San Diego, California 02182-1233, USA
  \and Departamento de F\'{\i}sica At\'omica, Molecular y Nuclear and
  Instituto Carlos I de F{\'\i}sica Te\'orica y Computacional, \\
  Universidad de Granada E-18071 Granada, Spain.}
\date{\today, Prepared for the special issue of "The tower of effective (field) theories and the emergence of nuclear phenomena''}
\abstract{
Are chiral theories at present describing experimental NN scattering data
satisfactorily ?. Will the chiral approach offer a framework where fitting and
selecting the existing np and pp data can be done without theoretical bias ?.
While predictive power in theoretical nuclear physics has been a major concern
in the study of nuclear structure and reactions, the Effective Field Theory
(EFT) based on chiral expansions has emerged after Weinberg as a model
independent hierarchy for many body forces and much progress has been achieved
over the last decades. We review some of the issues involved which point to
being close to the solution, but also that work remains still to be done to
validate the theory. We analyze several examples including zero energy NN
scattering and perturbative counter-term-free peripheral scattering where one
would expect these methods to work best and unveil relevant systematic
discrepancies when a fair comparison to the Granada-2013 NN-database and
partial wave analysis (PWA) based on coarse graining the interaction is
undertaken.
\PACS{
      {PACS-key}{describing text of that key}   \and
      {PACS-key}{describing text of that key}
     } % end of PACS codes
} %end of abstract
\maketitle
\section{Introduction}
\label{intro}

An old problem in Nuclear Physics concerns the predictive power of the theory,
which is persistently much poorer than experiment. For the compiled nuclear
masses one has typically $\Delta M (Z,N)^{\rm exp} < 1 {\rm
KeV}$~\cite{Audi:2014eak}. On the other hand the semi-empirical mass formula,
despite being an ancient and simple model with the liquid drop model
picture produces a much larger error of $\Delta M (Z,N)^{\rm th} $.
Sophisticated improvements based on mean field calculations have
achieved a benchmark $\Delta M (Z,N)^{\rm th} \sim 0.5 {\rm MeV} $ with a
large number of parameters~\cite{Goriely:2016gso}. After all the huge progress
made in recent years in the solution of the nuclear few- and many-body problem
alongside with the increasing computational power, a pending and open question
remains: can this predictive power be improved by truly acknowledging all
sources of uncertainties in a model independent fashion or at least
incorporating some true features of Quantum Chromodynamics (QCD)?.

Of course, the best possible answer is to solve QCD directly in terms of its
elementary degrees of freedom , quarks and gluons. Although much progress has
been made in lattice QCD calculations with respect to the nuclear problem (see
e.g. Refs.~\cite{Aoki:2011ep,Aoki:2013tba,Aoki:2012xa} for nuclear potential
studies), we are still not as accurate when compared to more
phenomenological approaches. One abusively refers usually to {\it ab initio}
calculations to determine atomic nuclei properties in terms of their
constituent nucleons. Nonetheless, there are QCD features such as chiral
symmetry which may be implemented in nuclear calculations. The above question
on the predictive power in Nuclear Physics still holds even if the specific
constraints on chiral symmetry are explicitly taken into account.

Chiral perturbation theory for the lightest $u$ and $d$ quarks is
based on the smallness of the pion mass as compared to the rest of
hadronic states such as the $\rho$-meson. Indeed, the existence of a
mass gap suggests that it should be possible to design an effective
Hilbert space where the dynamical degrees of freedom are just
pions. In addition, the fact that pions are the would-be Goldstone
bosons of the spontaneously broken chiral symmetry of QCD implies that
they couple derivatively, and hence they interact weakly at low
momenta~\cite{Weinberg:1968de}. This viewpoint together with the
general EFT idea~\cite{Weinberg:1978kz} can and has been efficiently
incorporated in the simplest $\pi\pi$ system and many successes have
followed~\cite{Gasser:1983yg}.  But similarly to nuclear physics,
hadronic interactions are usually characterized by a recurrent lack of
predictive power on the theory side as compared to the
experiments. The only known exception to this undesirable state of
affairs corresponds to the theoretical determination of $\pi\pi$
scattering lengths where the theory provides an estimate which is
about an order of magnitude more precise than the
experiment~\cite{Colangelo:2001df}. This has been possible thanks to
the EFT idea complemented with other properties. It is partly this
spectacular success which may be taken as a strong motivation to
incorporate and adapt these ideas elsewhere not only in Hadronic
Physics but also in Theoretical Nuclear Physics.

When Nucleons enter the game things become more difficult and subtle from
a theoretical perspective and the level of predictive power becomes more
compromised because the chiral expansion converges slowly even after the
regularization scheme in loop integrals is conveniently
designed~\cite{Gasser:1987rb,Bernard:1992qa,Becher:1999he}.  From a
phenomenological point of view, in Nuclear Physics the main practical reason
why chiral symmetry is not dominating could be found in the characteristic
small binding energies of nucleons in atomic nuclei, $B/A= 8 {\rm MeV}$ or
nuclear matter $B/A = 16 {\rm MeV}$ compared to typical hadronic scales
(including the pion mass, $m_\pi = 140 {\rm MeV}$) and therefore the
significance and impact of chiral symmetry is much more subtle and
questionable.

\begin{figure}[htb]
\begin{center}
\includegraphics[width = 0.95\linewidth]{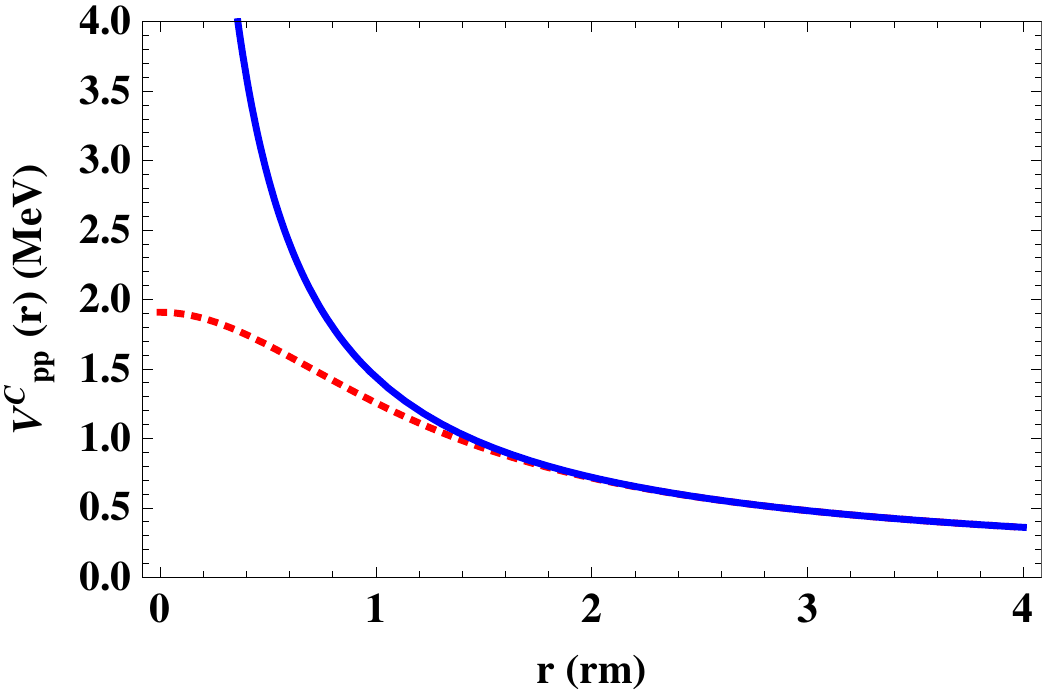}
\includegraphics[width = 0.95\linewidth]{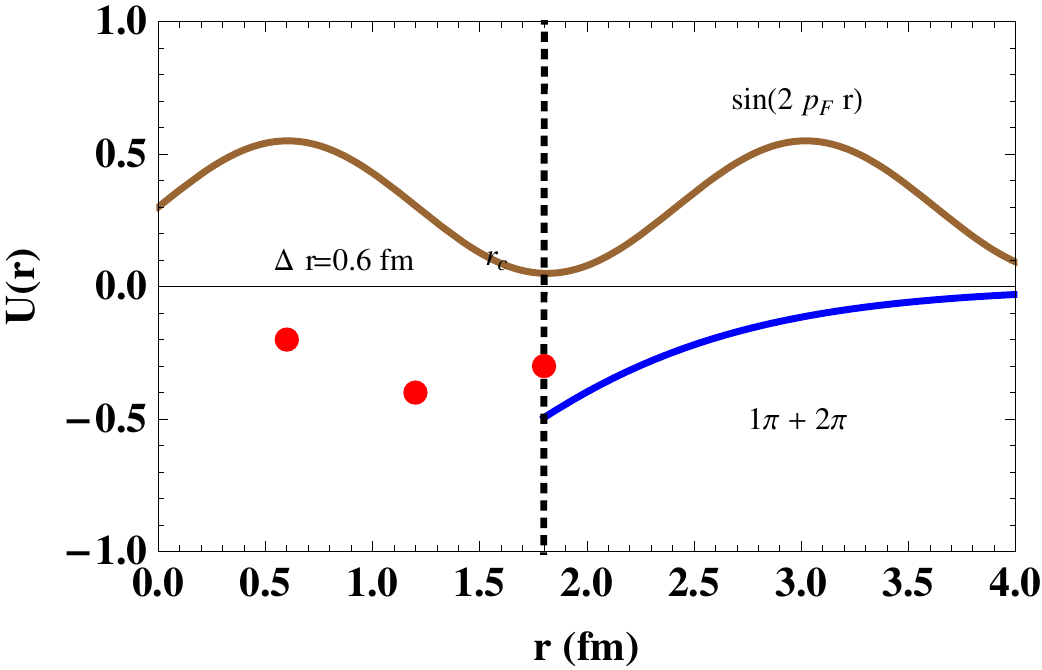}
\end{center}
\caption{\label{Fig:VU-r} Top panel: Point-like (solid-blue) and extended
(dashed-red) proton-proton Coulomb interaction as a function of distance.
Bottom panel: Coarse grained NN potential (red dots) plus the chiral $1\pi+2
\pi$ exchange potentials when $r_c=1.8$fm (blue line) compared with the wave
function at a CM corresponding to a back-to-back NN collision on the Fermi
surface in nuclear matter (brown line).}
\end{figure}

From the nucleon {\it ab initio} and reductionist perspective
the theoretical predictive power flow is expected to be from light to heavy
nuclei. Thus, from a Hamiltonian with multi-nucleon forces
\begin{equation}
H(A) = T + V_{2N} + V_{3N} + V_{4N}+ \dots \, ,   
\end{equation}
one proceeds to solve the Schr\"odinger equation 
\begin{equation}
H(A)\Psi_n = E_n(A) \Psi_n \, .  
\end{equation}
In the absence of useful QCD-{\it ab initio} determinations,
phenomenological $V_{2N}$ interactions are {\it adjusted} to NN
scattering data and the deuteron, $^2$H ($A=2$) and $V_{3N}$ to $^3$H
and $^3$He ($A=3$), $V_{4N}$ to $^4$He (A=4) binding energies and so
on. Within this setup, the chiral EFT approach to Nuclear Physics,
originally pioneered by Weinberg in 1990~\cite{Weinberg:1990rz} (see
e.g. \cite{Bedaque:2002mn,Epelbaum:2008ga,Machleidt:2011zz,Hammer:2019poc}
for reviews) to nuclear forces provides a power counting in terms of
the pion weak decay constant $f_\pi= 92$MeV, with the feature of
systematically providing an appealing hierarchy
\begin{equation}
  V_{2N}^{\chi} \gg V_{3N}^{\chi} \gg V_{4N}^{\chi} \gg \dots  , . 
\end{equation}
where the irreducible contributions of a $V_{nN}^\chi$ potentials contain
$(n-1)\pi$ exchanges as the longest range contributions. Because the pion mass
is so small, chiral interactions are local and unambiguous at long distances
via $1\pi$,$2\pi$,$3\pi,\dots$ exchanges for relative distances above a short
distance cut-off $r_c$, $V^{n \pi} (r_c) \sim e^{- n r_c m_\pi }$. Thus, if we
take nuclear matter at saturation density $\rho=0.17 {\rm fm}^{-3}$, the
average internucleon distance is $d \equiv \rho^{-\frac13} = 1.8 {\rm fm}$. We
have $ d m_\pi \sim 1.3 $ and for $n\pi$ exchange by a rough suppression
factor $\xi^n$ with $\xi = e^{-m_\pi d} \sim 0.3$. Of course, nucleons have a
finite size and are therefore characterized by an {\it elementary} radius
$r_e$ above which they interact as if they were elementary point-like particles
characterized by local fields. For instance, for the Coulomb $pp$ interaction
we have $V_{pp} (r) \sim e^2 /r$ for $r \ge r_e \sim 1.8 {\rm fm}$, whereas
for smaller distance the overlap between protons screens the interaction due
to their finite extension characterized by form factors. The numerical
coincidence between $d \sim r_e$ makes a strong case on what may be deduced
ignoring specific details for smaller distances, since for larger distances
the nucleon dynamics via pion exchange can be computed explicitly using
$\chi$PT. In Fig.~\ref{Fig:VU-r} we provide a pictorial picture of the
discussion above.

In the simplest NN case chiral potentials are constructed in perturbation
theory, are universal and contain chiral constants $c_1,c_3,c_4, \dots$ which
can be related to $\pi N$
scattering~\cite{Bedaque:2002mn,Epelbaum:2008ga,Machleidt:2011zz}. At long
distances we have
\begin{equation}
V^{\chi}_{NN}(r) =  V^{\pi}_{NN}(r) + V^{2 \pi}_{NN}(r) + V^{3 \pi}_{NN}(r) + \dots  \quad r \gg r_c \, , 
\end{equation}   
whereas they become singular at short distances 
\begin{equation}
V^{\chi}_{NN}(r) =  \frac{a_1}{f_\pi^2 r^3} +  \frac{a_2}{f_\pi^4 r^5} + 
\frac{a_3}{f_\pi^6 r^7} +  \dots  \qquad r \ll r_c \, ,
\label{eq:chi-short}
\end{equation} 
and some regularization must be introduced in {\it any} practical calculation.
This feature is also in common with 3N and 4N forces. Thus the following
questions arise: What is the best theoretical accuracy we can get within
``reasonable'' cut-offs? What is a reasonable cut-off? Can the short
distance piece be organized as a power counting compatible with the chiral
expansion of the long distance piece?

There has been a huge effort in the last 30 years based on the seminal
work of Weinberg in theoretical nuclear physics. In essence it
consists on a chiral expansion of the NN potential rather than the NN
amplitude. The subject of the present work will be some reflections on
the validation/falsification of NN chiral potentials when compared to
existing scattering pp and np data. In the present paper we do not
discuss the inner consistency of the Weinberg's power counting, a
subject which has been going on unsettled for 30 years now (see
e.g. \cite{Valderrama:2019lhj} for a recent discussion), but rather
try to confront it with the existent NN scattering data and wonder how
far are we from the claim that chiral symmetry ``works'' in Nuclear
Physics and what is actually meant by such a statement. We will
discuss here the simplest NN case, but the issues are worrisome enough
to reconsider the whole approach.  Motivated by this intriguing
possibility we have paid dedicated attention in the last years to the
issue of NN uncertainties including also chiral
interactions~\cite{Perez:2014kpa,Perez:2014waa}.

%Here we focus on
%$V_{NN}^{2\pi}$, corresponding to chiral $2\pi$ exchange ($\chi$TPE)
%and elaborate on the prospectives of improving the predictive power in
%nuclear physics.

The topic of the present work has certainly to do with proper
assessment and evaluation of uncertainties of any sort and in
particular in the NN interaction and its implications. For instance,
some time ago we made a first and simple
estimate~\cite{NavarroPerez:2012vr,Perez:2012kt} of $\Delta B^{\rm
  th}/A$ $\sim 0.5 \, {\rm MeV}$, which has been updated in
Ref.~\cite{Perez:2014waa} to be enlarged to $\sim 2 \, {\rm
  MeV}$. These crude estimates are in the bulk of a more recent
uncertainty analysis and order-by-order optimization of chiral nuclear
interactions~\cite{Carlsson:2015vda} including three-body forces.
These are large scale calculations where the numerical solution of the
nuclear problem is very much under control, and it is found $\Delta
B^{\rm th} ( ^{16}{\rm O}) /16 \sim 4 {\rm MeV}$. In fact, most of the
uncertainty is dominated by the cut-off variation within a
``reasonable'' range, but in any case it is much larger compared to
the ancient 5 parameter Weiszacker semi-empirical mass formula which
is based on the drop model picture, where one has $\Delta B_{\rm
  sem}/A \sim 0.1 {\rm MeV}$. One may thus be worried that the the
chiral approach to nuclear structure pioneered by Weinberg despite its
theoretical appeal might actually not be very accurate despite all the
new technical and conceptual sophistication which has followed
thereafter.

This work is partly of review character focusing mainly on work done
in Granada and adding some further aspects which have become clearer
in the analysis over the last years. For many details we will refer to
our previous publications along the present work. In our presentation
the interaction will be characterized by a conventional quantum
mechanical {\it potential}, which is not an observable itself (except
for static sources) similarly to the wave function. However, contrary
to what some people believe, besides being a convenient analysis tool
for NN scattering with a direct application to nuclear structure
calculations, its existence may be deduced from assumptions in quantum
field theory which are not more restrictive than those usually
made. In order to provide some scope we will review critically some of
the issues concerning the link between the scattering data and the
construction of the NN potential in general and the chiral potential
in particular.

\section{NN scattering}

\begin{figure}[htb]
\begin{center}
\includegraphics[width = 0.95\linewidth]{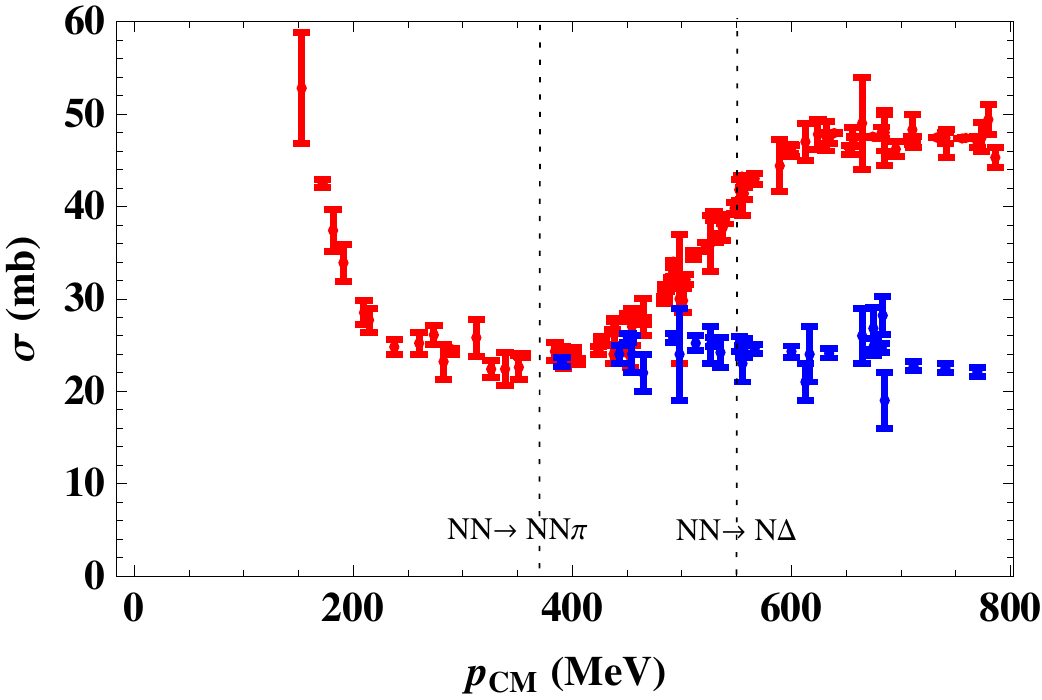}
\end{center}
\caption{\label{fig:sig-inel} The pp total (red) and total elastic (blue)
cross section as a function of the CM momentum. We mark the single
$\pi$-production and the $\Delta$-production for reference.}
\end{figure}

We will analyze pp and np scattering.  We assume for simplicity of
presentation the proton and neutron to have the same and common
nucleon mass $M_N$. At fixed LAB energy $T_{\rm LAB}= 2 k^2/M_N$, with
$k$ the CM momentum.  The pion production threshold $T_{\rm LAB}= 2
m_\pi + m_ \pi^2/ 2M_N \sim 290 $MeV i.e. $k=370$MeV, although it has
become customary to take $T_{\rm LAB} \le 350$MeV ($k \sim 400$MeV),
as the upper limit, since the inelastic cross section only becomes
comparable to the elastic between single- and double-$\Delta$
production, which corresponds to the range $T_{\rm LAB}= 650-1350$MeV
or equivalently $k \sim 550-800$MeV, see Fig.~\ref{fig:sig-inel}.
Thus, for $T_{\rm LAB} \le 350$MeV ( $p_{\rm CM} \le 400$MeV ) we may
assume {\it elastic} scattering.

\subsection{The scattering amplitude}

The elastic scattering process is described by the scattering matrix
$S$ whose matrix elements in the CM system are defined as $S(
\mathbf{\hat k}_f, \mathbf{\hat k}_i) \equiv \langle \mathbf{\hat k}_f
| S | \mathbf{\hat k}_i \rangle$ with $\mathbf{\hat k}_i $ and
$\mathbf{\hat k}_f $ the incoming and outgoing directions
respectively.  The unitarity of the S-matrix implies the condition
\begin{equation}
  \delta^{(2)} ( \mathbf{\hat k} - \mathbf{\hat k}') {\bf 1}= \int d^2 \mathbf{\hat k}'' S( \mathbf{\hat k}, \mathbf{\hat k}'')
  S( \mathbf{\hat k}'', \mathbf{\hat k}')^\dagger  \, ,
\end{equation}
where $ S(\mathbf{\hat k}_f, \mathbf{\hat k}_i) $ is a matrix in
spin-isospin space, so that we have $S( \mathbf{\hat k}_i,
\mathbf{\hat k}_f)^\dagger \equiv \langle \mathbf{\hat k}_f |
S^\dagger | \mathbf{\hat k}_i \rangle $. The NN scattering problem is
analyzed in terms of the scattering amplitude $M$, defined as,
\begin{equation}
  S( \mathbf{\hat k}_f, \mathbf{\hat k}_i )= {\bf 1} + 2 k i
  M(\mathbf{\hat k}_f,\mathbf{\hat k}_i) \, .
\end{equation}
If we impose P (Parity),T (Time Reversal) and Lorentz invariance
symmetries, the complete on-shell NN scattering amplitude contains
five independent complex quantities, which we choose for definiteness
as the Wolfenstein parameters~\cite{glockle1983quantum}
\begin{eqnarray}
M(\mathbf{\hat k}_f,\mathbf{\hat k}_i) &=& a + m (\mathbf{\sigma}_1 \cdot \mathbf{n})(\mathbf{\sigma}_2 \cdot \mathbf{n}) 
                  + (g-h)(\mathbf{\sigma}_1 \cdot \mathbf{m})(\mathbf{\sigma}_2,\mathbf{m}) \nonumber \\
                  & &+ (g+h)(\mathbf{\sigma}_1 \cdot \mathbf{l})(\mathbf{\sigma}_2 \cdot \mathbf{l})  + c(\mathbf{\sigma}_1+\mathbf{\sigma}_2) \cdot \mathbf{n} \, ,  
\end{eqnarray}
where $a,m,g,h,c$ depend on energy and angle, $\mathbf{\sigma}_1$ and
$\mathbf{\sigma}_2$ are the single nucleon Pauli matrices,
$\mathbf{l}$, $\mathbf{m}$, $\mathbf{n}$ are three unitary orthogonal
vectors along the directions of $ \mathbf{\hat k}_f+\mathbf{\hat k}_i
$, $ \mathbf{\hat k}_f-\mathbf{\hat k}_i $ and $ \mathbf{\hat k}_f
\wedge \mathbf{\hat k}_i $ and $\mathbf{k}_f = k \mathbf{\hat k}_f $,
$\mathbf{k}_i =k \mathbf{\hat k}_i $ are the final and initial
relative nucleon momenta respectively. The amplitudes $a,m,g,h,c$
could in principle be determined directly from experiment as shown in
Ref.~\cite{Hoshizaki:1969qt,Bystricky:1976jr,LaFrance:1981bg} (see
also ~\cite{kamada2011} for an exact analytical inversion). In
writing this amplitude, use has been made of the on-shell elastic
condition $ k_f = k_i $ which implies the
identity~\cite{okubo1958velocity},
\begin{equation}
\mathbf{\sigma_1} \cdot \mathbf{\sigma_2} =
  (\mathbf{\sigma}_1 \cdot \mathbf{l})(\mathbf{\sigma}_2 \cdot \mathbf{l}) +
  (\mathbf{\sigma}_1 \cdot \mathbf{m})(\mathbf{\sigma}_2 \cdot \mathbf{m}) +
(\mathbf{\sigma}_1 \cdot \mathbf{n})(\mathbf{\sigma}_2 \cdot \mathbf{n})
\label{eq:on-id}
\end{equation}
As a consequence of unitarity we have the relation
\begin{equation}
  M(\mathbf{\hat k}_f,\mathbf{\hat k}_i) - M(\mathbf{\hat k}_f,\mathbf{\hat k}_i)^\dagger
  =  2 i k \int d^2 \hat k_n M(\mathbf{\hat k}_f,\mathbf{\hat k}_n) 
  M(\mathbf{\hat k}_n,\mathbf{\hat k}_i)^\dagger  \, . 
\end{equation}  
If we go to the orthonormal basis of eigenstates $\phi_n(\mathbf{ \hat k})$
with eigenvalues $M_n$ the spectral decomposition reads 
\begin{equation}
M(\mathbf{\hat k}_f,\mathbf{\hat k}_i) = \sum_n M_n \phi_n(\mathbf{ \hat k}_f)
\phi_n(\mathbf{ \hat k}_i)^\dagger  \, , 
\end{equation}
so that we can write the unitarity condition as $M_n = K_n /(1 -ik K_n)$
with $K_n$ real, and define the self-adjoint operator 
\begin{equation}
K(\mathbf{ \hat k}_f,\mathbf{ \hat k}_i) = \sum_n K_n \phi_n(\mathbf{\hat k}_f)
\phi_n(\mathbf{ \hat k}_i)^\dagger  
\end{equation}
and we have the integral equation
\begin{equation}
  M(\mathbf{\hat k}_f,\mathbf{\hat k}_i) = K(\mathbf{\hat k}_f,\mathbf{\hat k}_i) +
  i k \int d^2 \hat k_n K(\mathbf{\hat k}_f,\mathbf{\hat k}_n) M(\mathbf{\hat k}_n,\mathbf{\hat k}_i) \, . 
  \label{eq:int-on}
 \end{equation}
The $K-$matrix has the same decomposition as the scattering amplitude
but now the coefficients are {\it real} which means that scattering at
a given energy and angle can be described by 5 independent real
functions.

\subsection{Analytical properties}

To these properties, one has to add the existence of ana\-ly\-ti\-cal
properties in the scattering energy and angle, or equivalently in the
Mandelstam variables $t= -(\mathbf{k}_f-\mathbf{k}_i)^2$ and $s= 4 (k^2 +
M_N^2)$ in the complex plane. This corresponds to a Mandelstam double spectral
representation of the scattering amplitude~\cite{Mandelstam:1959bc} which
actually provides a justification for using interpolation methods in
phenomenological analyses. These important constraints are explicitly
satisfied in field theory in perturbation theory, where interactions arise
from particle exchange and at long distances pion exchanges dominate.
Unfortunately, the unitarity of the S-matrix is not preserved {\it exactly} in
perturbation theory and several unitarization methods have been proposed.
While all these elements provide a framework to describe the scattering
problem it does not give a direct hint about the NN interactions from which
nuclear binding energies might be determined. It should be noted that for
quantum mechanical potentials being a superposition of Yukawa potentials the
double spectral representation holds for elastic
scattering~\cite{Blankenbecler:1960zz} and also in the presence of
inelasticities described by a complex and energy dependent optical
potential~\cite{cornwall1962mandelstam,omnes1965optical}. This issue has been
exemplified recently in the case of
$\pi\pi$-scattering~\cite{RuizdeElvira:2018hsv}.

\subsection{The partial wave expansion}

The NN scattering amplitude conserves the total angular momentum $\vec J= \vec
L+ \vec S$, the spin $(\vec S)^2$, and hence a complete set of commuting
observables is given by $\{J^2,J_z,S^2 \}$ so that a convenient basis is given
by the vector spherical harmonics ${\cal Y}_{JLSM} (\hat k)$, so that
\begin{equation}
S \,{\cal Y}_{JLSM} (\hat k) = \sum_{L'} S^{J,S}_{L,L'} {\cal Y}_{JL'SM} (\hat k).
\end{equation}
The analysis of NN scattering has been traditionally carried out by a
decomposition of the scattering amplitude in partial waves. For this amplitude
the partial wave expansion in this case reads
\begin{align}
 M^s_{m_s',m_s}(\theta) &=  \frac{1}{2ik} \sum_{J,l',l}\sqrt{4\pi(2l+1)}Y^{l'}_{m_s'-m_s}(\theta,0) \nonumber \\
      &\times C^{l',S,J}_{m_s-m_s',m_s',m_s}i^{l-l'}(S^{J,S}_{l,l'}-\delta_{l',l}) C^{l,S,J}_{0,m_s,m_s}, 
 \label{eq:MmatrixPartialWaves}
\end{align}
where $S$ is the unitary coupled channel S-matrix, and the $C's$ are
Clebsch-Gordan coefficients. Denoting the phase shifts as
$\delta^{J,s}_{l,l'}$, for the singlet ($s=0$, $l = l'= J$) and triplet
uncoupled ($s=1$, $l=l'=J$) channels the $S$ matrix is simply
$e^{2i\delta^{J,s}_{l,l}}$, in the triplet coupled channel ($s=1$, $l=J\pm1$,
$l'=J\pm1$) it reads
\begin{equation}
S^J = \left( 
\begin{array}{c c}
 e^{2i\delta^{J,1}_{J-1}} \cos{2 \epsilon_J} & ie^{i(\delta^{J,1}_{J-1}+\delta^{J,1}_{J+1})} \sin{2 \epsilon_J} \\
 ie^{i(\delta^{J,1}_{J-1}+\delta^{J,1}_{J+1})} \sin{2 \epsilon_J} & e^{2i\delta^{J,1}_{J+1}} \cos{2 \epsilon_J}
\end{array}
\right),
\end{equation}
with $\epsilon_J$ the mixing angle.

Because of unitarity one has that ${\bf S}^{JS} = ({\bf M}^{JS} - i {\bf
1})({\bf M}^{JS} + i {\bf 1})^{-1} $ with $({\bf M}^{JS})^\dagger = {\bf
M}^{JS}$ a hermitian coupled channel matrix (also known as the K-matrix). The
main advantage is that for finite range interactions of range $a$ one expects
the partial wave sum to be truncated at about $L_{\rm max} +1/2 \sim k a $.

\begin{figure*}[htb]
\begin{center}
\includegraphics[width=0.95\linewidth]{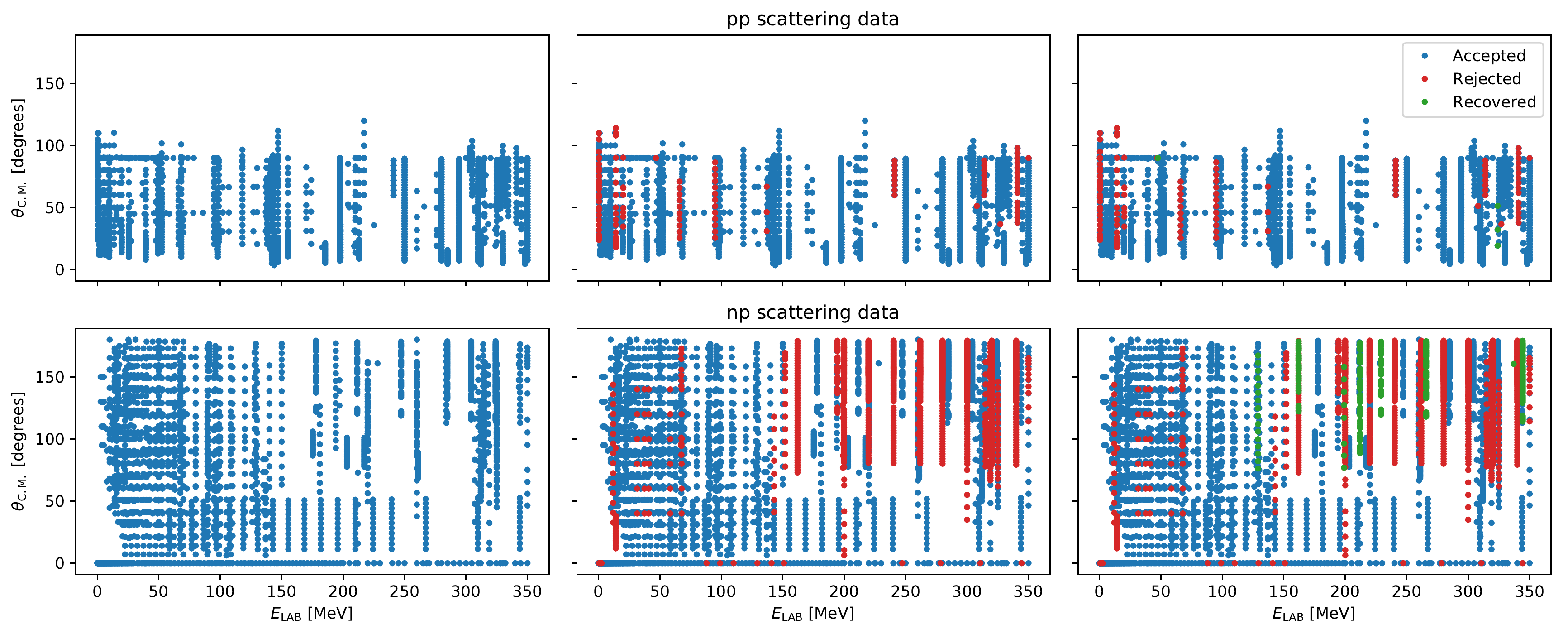}
\end{center}
\caption{\label{fig:NN-abundance} Abundance plots for pp (top panels) and np
(bottom panels) scattering data. Full data base (left panel). Standard
$3\sigma$ criterion (middle panels). Self-consistent $3\sigma$ criterion (right
panels). We show accepted data (blue), rejected data (red) and recovered data
(green).}
\end{figure*}

As mentioned, the NN scattering amplitude has 5 independent complex components
for any given energy, which must and can be determined from a complete set of
measurements involving differential cross sections and polarization
observables. While this is most often done in terms of phase-shifts, it is
worth reminding that phase shifts obtained in PWA are {\it not} observables by
themselves. This is so unless a complete set of 10 fixed energy and angle
dependent measurements have been carried out. This is a rare case among the
bunch of existing 8000 np+pp scattering data below $350 \, {\rm MeV}$ LAB
energy and which corresponds to a maximal CM momentum of $p_{\rm CM}^{\rm
max}= 2 {\rm fm}^{-1}$. In order to intertwine all available, often incomplete
and partially self-contradictory, information some energy interpolation is
needed. The situation is illustrated by the abundance plots in 
Fig.~\ref{fig:NN-abundance} (see top and bottom left panels) where every point
represents a measured observable (cross section, polarization, etc.) of a
total of about 8000 pp+np data.  Thus, the fact that the energy dependence of
the amplitude is not completely arbitrary will be most helpful.

\begin{figure}[htb]
\begin{center}
\includegraphics[width = 0.95\linewidth]{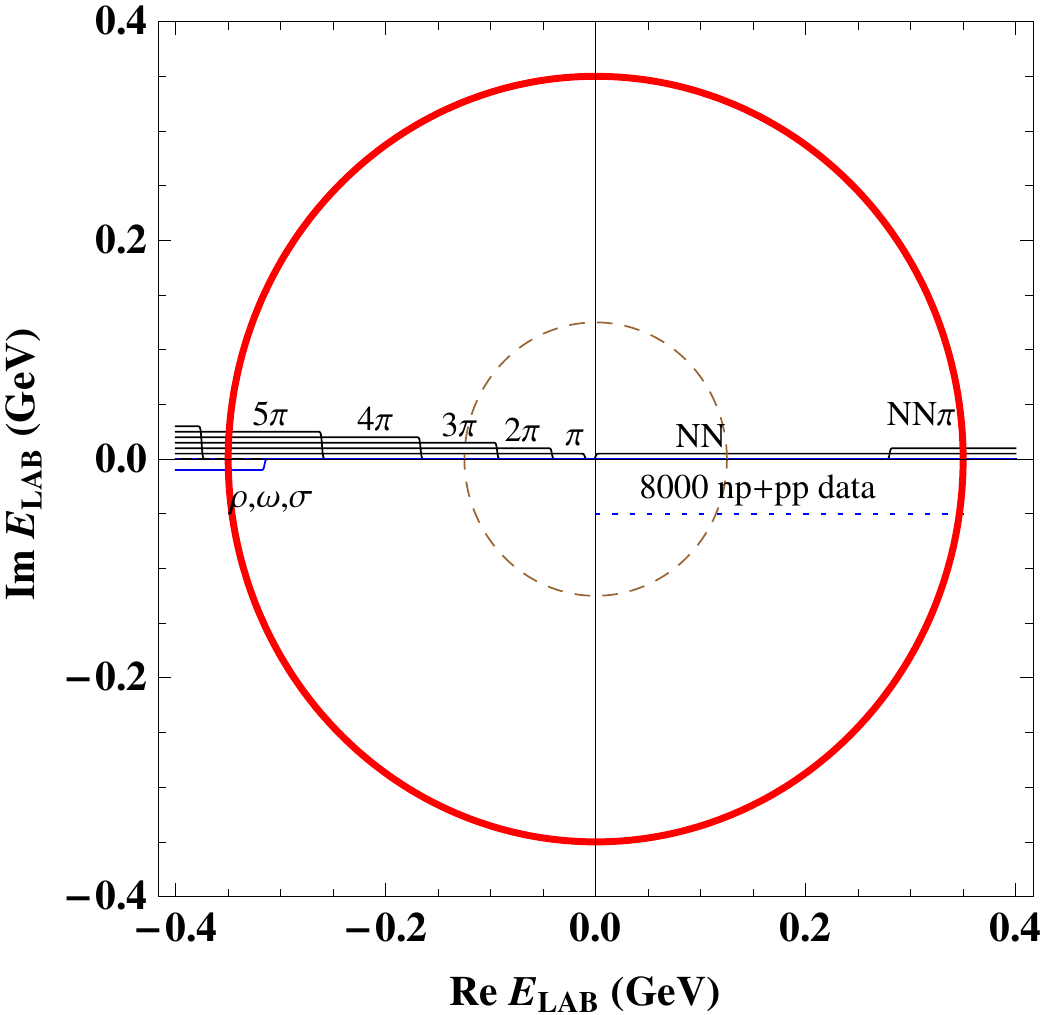}
\end{center}
\caption{\label{Fig:TLAB-complex} The LAB energy complex plane, showing the
partial waves left cut structure due to multiple pion (and
$\sigma,\rho\omega$) exchange along with the right cut structure due to pion
production. $T_{\rm LAB}^{\rm left}= ( \dots,
-375.3,-260.6,-166.8,-93.8,-41.7,-10.4) {\rm MeV}$. The outer/inner circles
correspond to LAB energies of $350/125$MeV respectively.}
\end{figure}

At the level of partial waves the multi-pion exchange diagrams generate left
hand cuts in the complex s-plane, which come in addition to the NN elastic
right cut and the $\pi NN$, $ 2\pi NN$ etc., pion production cuts,as can be
seen in Fig.~\ref{Fig:TLAB-complex}. At low energies for $ |p| \le m_\pi/2$
the scattering amplitude is analytic and we have the Taylor
expansion~\cite{PavonValderrama:2005ku}
\begin{equation}
  p^{l+l'+1} M_{l,l'}^{JS} (p) = -(\alpha^{-1})^{JS}_{l,l'} + \frac12  (r_0)^{JS}_{l,l'}p^2+ \sum_{n=2}^\infty (v_{2n})^{JS}_{l,l'}p^{2n} 
\label{eq:ERE-coup}
\end{equation}
An implementation of these analytical properties can be done in terms
of dispersion relations for short range interactions and thus leaving
out the important case of long range interactions such as Coulomb and
magnetic moments interactions. In Fig.~\ref{Fig:TLAB-complex} we
depict a characteristic contour which in fairness would be needed for
encompassing the available data with $T_{\rm LAB} \le 350$MeV along
the unitarity cut but also up to about $5\pi$-exchange to faithfully
describe the left hand cut within the same contour. At present only
$3\pi$ exchanges have been considered starting by
Ref.~\cite{Pupin:1999ba}. A particularly attractive scheme to
represent the analytical properties is given by the so-called N/D
method, where the partial wave amplitude is represented as a ratio
between two functions $N(s)$ which has only left-cut (particle
exchange) discontinuities and $D(s)$ which has only right-cut
(unitarity) discontinuities. While this method has been around for
over 50 years, in the NN case the resulting set of integral equations
required from multi-pion exchange are highly singular at short
distances and only a suitable subtraction hierarchy allows to handle
the singularities (see e.g. \cite{Oller:2018zts} and references
therein). These conclusions have a parallel developement in terms of a
quantum mechanical potential for the renormalization in coordinate
space~\cite{PavonValderrama:2005wv,PavonValderrama:2005uj} and the
$N/D$ representation~\cite{RuizdeElvira:2018hsv}.

\section{The NN potential}

\subsection{The use of a potential}

One of the good reasons to discuss the NN scattering problem is to design
suitable NN interactions which can be used in few and many body calculations
and in particular address the problem of binding in atomic nuclei. In the most
popular Hamiltonian approach the interaction is characterized by a {\it
potential}. Moreover, provided the potential fulfills a spectral
representation as a superposition of Yukawa One Pion Exchange (OPE) form, the
scattering amplitude can be shown to posses analytical properties in both the
CM energy and the scattering angle (or equivalently Mandelstam $s,t$
variables), which also guarantees the smoothness in a given energy
interpolation which proves useful in fitting data.

For LAB energies below $350$MeV relativity plays a small but significant role
(relativistic Coulomb corrections will be needed). The proper incorporation of
relativity requires also to take into account retardation effects. The
standard Bethe-Salpeter equation~\cite{Salpeter:1951sz} is an exact
four-dimensional integral equation which besides facing technical difficulties
poses theoretical issues in practice since any finite truncation of
irreducible diagrams generates spurious effects and inconsistencies in the
amplitude or generate fake results in the heavy-light limiting case. One may
consider instead three dimensional reductions of the
Blankenblecker-Sugar~\cite{Blankenbecler:1965gx}, Gross~\cite{Gross:1969rv} or
Kadyshevsky form \cite{Kadyshevsky:1967rs}, among the many possible schemes,
fulfilling special properties. This relativistic ambiguity would come in
addition to several ones discussed below.

Assuming from now on the non-relativistic case, NN scattering is
formulated in terms of the Lippmann-Schwinger (LS) equation which at the
operator level reads
\begin{equation}
  T(E) = V + V G_0 T(E) \, ,
  \label{eq:LS}
\end{equation}
where $H_0$ is the free Hamiltonian and $V$ is the potential and $E=
k^2 /2 \mu $ is the CM energy and $G_0=(E-H_0)^{-1}$. Inserting a
complete set of states we have
\begin{align}
  \langle \vecp' | T(E) | \vecp \rangle &=    \langle \vecp' | V | \vecp \rangle \label{eq:LS-f}  \\ 
  &+ \int \frac{d^3 p''}{(2\pi)^3} \langle \vecp' | V | \vecp'' \rangle \frac{1}{E-E_{p''}}
  \langle \vecp'' | T(E) | \vecp \rangle  \, . \nonumber
\end{align}
Then, taking the on-shell conditions, we get 
\begin{equation}
M (\hat \vecp', \hat \vecp )= - \frac{\mu}{2 \pi }\langle \vecp' | T (E) | \vecp \rangle |_{|p'|=|p|=k} \, . 
\end{equation}
Note that although Eq.~(\ref{eq:int-on}) and Eq.(\ref{eq:LS-f}) look
very similar they are in fact rather different since the second
equation contains in addition an integral over the energy, and thus
the relation between $K$ and $V$ is an integral equation. The LS
framework generates therefore an ambiguity in the potential, i.e.
there are infinitely many potentials which generate the same on-shell
scattering amplitude. While we may use this freedom to impose certain
convenient conditions, one must not abuse this possibility since it
may introduce a bias in the analysis of the data. Being ourselves
theoreticians, we take the practical point of view that the least
biased choice of the interaction corresponds to the one allowing to
congregate as many data as possible, rather than pondering on the
correctness of the many experiments.

From a purely mathematical perspective, the inverse scattering problem
concerns the determination of such a potential {\it directly} from the
scattering data~\cite{chadan2012inverse}. As such, the solution of this
problem is ambiguous and additional conditions need to be imposed to fix this
ambiguity. In practice, implementing this approach requires a parameterization
of the scattering amplitude interpolating between different measured points,
which becomes rather involved and may require a large number of interpolating
parameters if high precision is required\footnote{This is the case for
instance in the NN SAID analysis~\cite{Workman:2016ysf} and several other
papers where the discrepancy of the phases for a large number of parameters is
larger than the statistical uncertainty (see e.g. \cite{Kohlhoff:1994gd}).}.

As we will review below, the scheme which is usually employed is to
make instead a least squares determination of a {\it proposed} form of
the potential in terms of the measured scattering observables which
may be validated or falsified in a statistical sense. Such an approach
has allowed to describe about 7000 np+pp scattering measurements below
$T_{\rm LAB}=350 {\rm MeV}$ with a number of about 30-50 parameters
which a high degree of confidence. Of course, in the case that a given
NN potential is validated, error analysis applies and experimental
errors can be propagated to any predicted quantity.

\subsection{NN potential components}

Assuming isospin invariance, the most general form of the NN
interaction can be written as \cite{okubo1958velocity}
\begin{align} 
V({\vecp}~', \vecp) &  = 
 \:\, V_C \:\, + \btau_1 \cdot \btau_2 \, W_C 
\nonumber \\ &+   
\left[ \, V_S \:\, + \btau_1 \cdot \btau_2 \, W_S 
\,\:\, \right] \,
\vec\sigma_1 \cdot \vec \sigma_2
\nonumber \\ &+ 
\left[ \, V_{LS} + \btau_1 \cdot \btau_2 \, W_{LS}    
\right] \,
\left(-i \vec S \cdot (\vecq \times \vecP) \,\right)
\nonumber \\ &+ 
\left[ \, V_T \:\,     + \btau_1 \cdot \btau_2 \, W_T 
\,\:\, \right] \,
\vec \sigma_1 \cdot \vecq \,\, \vec \sigma_2 \cdot \vecq  
\nonumber \\ &+ 
\left[ \, V_{Q} + \btau_1 \cdot \btau_2 \, 
      W_{Q} \, \right] \,
\vec\sigma_1\cdot(\vecq\times \vecP\,) \,\,
\vec \sigma_2 \cdot(\vecq\times \vecP\,)
\nonumber \\ &+ 
\left[ \, V_{P} + \btau_1 \cdot \btau_2 \, 
      W_{P} \, \right] \,
\vec\sigma_1\cdot \vecP \,\,
\vec \sigma_2 \cdot \vec  P\,
%\bigg\}
\, ,
%\nonumber \\ && 
\label{eq_nnamp}
\end{align}
where ${\vecp}\,'$ and $\vecp$ denote the final and initial nucleon
momenta in the CMS, respectively. Moreover, $\vecq = {\vecp}\,' -
\vecp$ is the momentum transfer, $\vecP =({\vecp}\,' + \vecp)/2$
the average momentum, and $\vec S =(\vec\sigma_1+ \vec\sigma_2)/2 $
the total spin, with $\vec \sigma_{1,2}$ and $\btau_{1,2}$ the
spin and isospin operators, of nucleon 1 and 2, respectively.

For the on-shell situation, $V_i$ and $W_i$ (where $i$ is equal to $C$, $S$,
$LS$, $T$, $Q$ or $P$) can be expressed as functions of $q= |\vecq\,|$ and
$p=|{\vecp}\,'| = |\vecp\,|$, only. As pointed out in
Ref.~\cite{okubo1958velocity} the terms corresponding to $V_P$ and $W_P$ can
be re-written in terms of other operators {\it provided} particles are
on-shell, i.e. $ \vecP \cdot \vecq = p'^2-p^2=0$. However, this is in general not the
case. For instance the exchange of an $A_1$ meson does produce such a
contribution to order $1/M_N^2$.

The calculation of the potential stemming from field theory always proceeds by
matching the perturbative solution of the quantum mechanical problem to a
perturbative calculation of Feynman diagrams in quantum field
theory~\cite{Logunov:1963yc}. For instance, the leading contribution in our
sign-convention is such that the one-pion exchange contribution in is of the
form $W_T^{(1\pi)} = - (g_{\pi N}/2M_N)^2 (m_\pi^2+q^2)^{-1}$, with the
physical values of the coupling constant $g_{\pi N}$ and the nucleon and pion
masses $M_N$ and $m_\pi$. In general, for a quantum field theory (QFT)
calculation organized in the perturbative expansion
\begin{equation}
T|_{\rm QFT} = T_1 + T_2 + \cdots
\end{equation}
we assume a similar expansion of the potential
\begin{equation}
 V = V_1 + V_2 +\cdots  
\end{equation}
such as the perturbative solution of the LS equation, Eq.~(\ref{eq:LS})
is identified order by order
\begin{align}
  V_1 &= T_1  \nonumber \\
  V_2 &= T_2 - V_1 G_0 V_1  \nonumber \\
    V_3 &= T_3 + V_1 G_0 V_1  G_0 V_1  - V_2 G_0 V_1 - V_1 G_0 V_2  \nonumber \\ 
  \dots
\end{align}
This procedure introduces ambiguities, since strictly speaking only
the resulting on-shell scattering amplitude is u\-niquely
defined. Therefore, there is no unique way of determining the
potential, even in perturbation theory, a perturbative reminiscent
feature from the inverse scattering problem. The remaining freedom can
be used advantageously to {\it choose} a particular form of the
potential by means of suitable unitary
transformations~\cite{Amghar:2002pf}. The advantage is that the
potential may be tailored to be used within a given computational
scheme solving the nuclear problem for finite nuclei. In a broader
context one should, however, keep in mind that this has also some
implication on the 3N problem, since the very definition of a
three-body force is based on the definition of the two-body
interaction.

As it is well known, unitarity is not preserved {\it exactly} in perturbation
theory, only order by order. Therefore, the identification of scattering
quantities, say the phase-shifts, is not unique.  From this point of view one
of the motivations to proceed via the potential, is that unitarity is restored
at any order of the calculation, since the total amplitude corresponds to a
particular re-summation method. Again, the unitarization method is not unique,
and different re-summation schemes yield different results (see e.g.
Ref.~\cite{Nieves:1999bx} for a discussion in the $\pi\pi$ case). A good
example is the choice of the 3D reduction of a relativistic scattering
equation for which several possibilities exist.

Moreover, perturbation theory is based on the smallness of a coupling
constant; the amplitude is parametrically small when the coupling
constant tends to zero. The real convergence for a finite coupling
constant is another issue, and one may wonder under what physical
conditions does a perturbative calculation provide sensible
results. Naively, we should expect small angles scattering, or
equivalently $q \to 0$, to be the relevant situation. In practice the
finite pion mass, requires the unphysical limit $q \to i m_\pi$ for
the OPE contribution to dominate over the other contributions, but
there is no measurable kinematic region where for instance the
coupling constant can be determined from the OPE potential {\it
  alone}; either some extrapolation from the data into the unphysical
region is needed or a short range contribution must be included to fit
experimental data. At present the most accurate determination to data
uses the second method~\cite{Perez:2016aol,Arriola:2016hfi}. The small
angle limit corresponds at the partial waves level to large angular
momentum states and in the classical limit to a large impact
parameter.  These states are not measured directly but rather deduced
from a PWA. We will comment on this issue below in
more detail when discussing peripheral waves.

\subsection{Semi-local Potentials}

The scalar functions appearing in the potential, Eq.~(\ref{eq_nnamp}),
depend on {\it both} initial and final momentum $\vec{p}$ and $\vecp'$
respectively. Because of rotational invariance we may thus form three
independent invariants, such as $p,p'$ and also $\vecq \cdot \vecP$
(which vanishes on-shell). Passing to coordinate space in the
conjugate variable of $q$ we have
\begin{equation}
  V(\vecr, \vecP) = \int \frac{d^3 q}{(2\pi)^3}
  e^{i \vecq \cdot \vecr } \langle \vecP + \frac12 \vecq | V | \vecP - \frac12 \vecq \rangle \, , 
\end{equation}
where we take $\langle \vecP + \frac12 \vecq | V | \vecP - \frac12
\vecq \rangle \equiv V(\vecp',\vecp)$.  The case where these functions
depend {\it only} on the momentum transfer $\vecq=\vecp'-\vecp$
corresponds in coordinate space to a {\it local} potential,
$V(\vecr,\vecP)= V(\vecr)$. This has the appealing property that the
quantum mechanical problem becomes a differential equation which can
be solved very efficiently after imposing regularity conditions of the
wave function at the origin. Moreover, important long-range effects
such as the Coulomb interaction or the magnetic moments interaction
are local and the incorporation in coordinate space is straightforward
and much less painful than in momentum space. However, {\it imposing}
this particular form is a restriction which might introduce a bias in
the statistical analysis of the scattering data. In the general case,
the presence of polynomials in $P$ implies that the potential is also
a differential operator, as can be checked by taking the corresponding
Fourier transformation.

There is a limiting case, however, where we expect the potential to be
truly local and actually an observable. Indeed, attaching a field
theoretical interpretation to the interaction, locality must be
satisfied by heavy and point-like elementary nucleons which act as
static sources. In this case, the static energy between them
corresponds to the potential
\begin{equation}
E_{NN}(r) = V_{NN} (r) + 2 M_N + {\cal O} ( M_N^{-1}) \, ,
\end{equation}
where we assume $M_N \gg m_\pi, E $.  Thus, we expect finite nucleon mass
effects to be responsible for non-localities, which means that we will have the
combination $P/M_N$. Thus, it makes sense to assume a polynomial in $P/M_N$,
which upon transformation into Fourier space will make the potential a
differential operator. This form has been frequently used in the past up to
${\cal O} ( P^2) $ because it still corresponds to a second order differential
equation, in the generalized Sturm-Liouville form with the standard regularity
conditions at the origin (see e.g. Ref.~\cite{Piarulli:2016vel})\footnote{One
can equivalently make a change of variables, both in the coordinate $r$ and in
the wave function $\Psi(r)$ so that the equation is effectively transformed
into a conventional Schr\"odinger potential.}. However, to order ${\cal O}
(P^4) $ or higher, one should impose regularity conditions and the wave
function and higher derivatives; this is perhaps the reason why to our
knowledge it has never been implemented. The momentum space approach, however,
does not require explicitly these boundary conditions at the origin, and this
is the preferred form in case of strong non-locality.

As already mentioned, it is possible, by means of suitable unitary
transformations~\cite{Amghar:2002pf} to transfer the operators $P$ into
angular momentum operators, at least to ${\cal O} ( P^2) $, which act
matrix-multiplicatively on the partial wave basis and hence no derivatives of
the wave function need to be evaluated. In practice, phenomenological and
chiral potentials may be taken to be weakly nonlocal or {\it semi-local}. The
computational advantages for a local interaction have been exploited in the
Argonne potentials saga which were specifically designed to favor MonteCarlo
calculations\footnote{One of the early arguments to favor this type of
interaction was the observation that non-localities such as those present in
the Paris potential would generate huge effects in nuclear
matter~\cite{Lagaris:1981mm}.}.

\begin{figure*}[htb]
\begin{center}
\includegraphics[width=\linewidth]{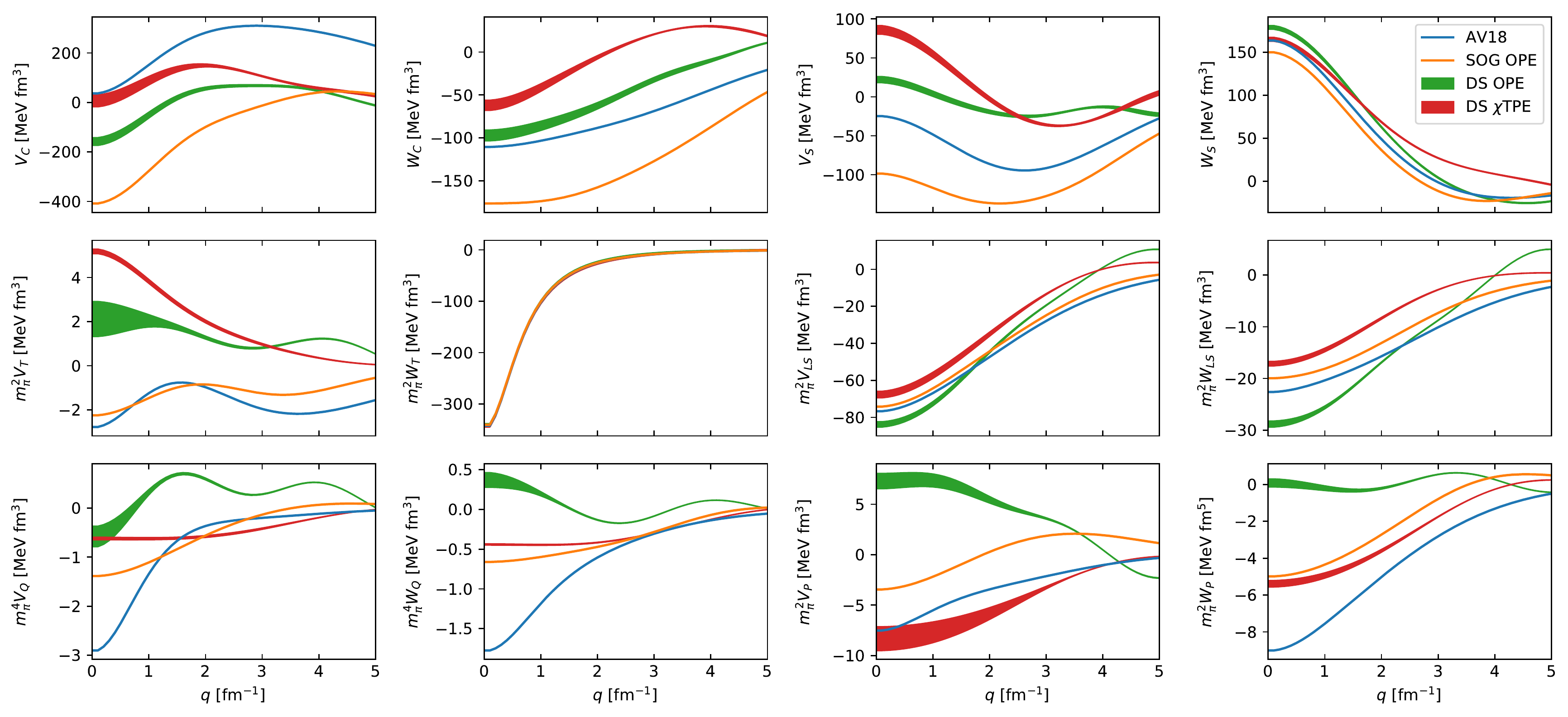}
\end{center}
\caption{\label{fig:mom-sys-local} Momentum space components of
  potentials as a function of the momentum transfer for several
  potentials: AV18~\cite{Wiringa:1994wb} and the Granada
  DS-OPE~\cite{Perez:2013mwa,Perez:2013jpa},
  DS-$\chi$TPE~\cite{Perez:2013oba,Perez:2013cza} and
  Gauss-OPE~\cite{Perez:2014yla}.}
\end{figure*}

In the Argonne basis, the potential containing at most two powers of momentum 
can be written as
\begin{equation}
V= \sum_{n=1}^{21} O_n   V_n (r) \, , 
\end{equation}
where the operators $O_n $ contain the linear momentum operator {\it only}
through the angular momentum operator $\vec L= \vecr \wedge \vecp$ and are
matrix-multiplicative when acting on the partial wave basis, i.e.
\begin{equation}
	O_n {\cal Y}_{J L S M} (\hat r) = \sum_{L'} (O_n)_{L,L'}^{JS} {\cal Y}_{J L' S M} (\hat r).
\end{equation}
This can be generalized to higher orders in $p$ or equivalently in $L$. Here,
the first fourteen operators are the same charge-independent ones used in the
Argonne $v_{14}$ potential and are given by
\begin{align}
   O^{n=1,14} &= 1,  {\tau}_{1}\!\cdot\! {\tau}_{2},\,
        {\sigma}_{1}\!\cdot\! {\sigma}_{2},
       ( {\sigma}_{1}\!\cdot\! {\sigma}_{2})
       ( {\tau}_{1}\!\cdot\! {\tau}_{2}),\,  S_{12}, S_{12}( {\tau}_{1}\!\cdot\! {\tau}_{2}),\, \nonumber \\ 
    &
    {\bf L}\!\cdot\!{\bf S}, {\bf L}\!\cdot\!{\bf S}
        ( {\tau}_{1}\!\cdot\! {\tau}_{2}), L^{2}, L^{2}( {\tau}_{1}\!\cdot\! {\tau}_{2}),\,
    L^{2}( {\sigma}_{1}\!\cdot\! {\sigma}_{2}), \nonumber\\
    & 
    L^{2}( {\sigma}_{1}\!\cdot\! {\sigma}_{2})
         ( {\tau}_{1}\!\cdot\! {\tau}_{2}),\, 
    ({\bf L}\!\cdot\!{\bf S})^{2}, ({\bf L}\!\cdot\!{\bf S})^{2}
        ( {\tau}_{1}\!\cdot\! {\tau}_{2})\ .
\end{align}
These fourteen components are denoted by the abbreviations $c$, $\tau$,
$\sigma$, $\sigma\tau$, $t$, $t\tau$, $ls$, $ls\tau$, $l2$, $l2\tau$,
$l2\sigma$, $l2\sigma\tau$, $ls2$, and $ls2\tau$. The remaining  terms are
\begin{align}
   O^{n=15,21}  &=  T_{12}, \,
        ( {\sigma}_{1}\!\cdot\! {\sigma}_{2})
       , T_{12}\, S_{12}T_{12},\, (\tau_{z1}+\tau_{z2})\ , \nonumber \\ 
&( {\sigma}_{1}\!\cdot\! {\sigma}_{2}) (\tau_{z1}+\tau_{z2})\ ,
L^{2} T_{12} ,  L^{2} ( {\sigma}_{1}\!\cdot\! {\sigma}_{2}) T_{12} \, . 
\end{align}
These terms are charge dependent and are labeled as $T$, $\sigma T$,$tT$,
$\sigma \tau z$, $l2T$ and $l2\sigma T$.

\subsection{Momentum space representation}
The relation between the Argonne basis and the momentum space representation
in Eq.~(\ref{eq_nnamp}) can be found passing to momentum space 
\begin{equation}
  \langle \vecp' | V | \vecp \rangle \equiv \int \frac{d^3
    r}{(2\pi)^3} e^{-i \vecp' \cdot \vec r} V e^{i \vecp \cdot \vec
    r} = \sum_n \tilde O_n (q,P) \tilde V_n (q)
\end{equation}
We list the basic Fourier transforms in the Appendix~\ref{sec:mom-sp}
and operator transforms (see e.g. Ref.~\cite{Veerasamy:2011ak}).  Two
aspects deserve attention, off-shellness and locality.  In general the
12 components depend on the scalars $q, \vecP \cdot \vecq, P$ and they
reduce to just 10 when the on-shell condition $\vecP \cdot \vecq=0 $ is
taken. The local {\it approximation} corresponds to take $\vecP=0$ which
includes also the on-shell condition. In Fig.~\ref{fig:mom-sys-local}
we depict the 12 components for the AV18
potential~\cite{Wiringa:1994wb} in the local approximation. We see
that while not all potentials are of the same size, all allowed
components are non-vanishing. This observation will be useful when
analyzing the implications of power counting and chiral symmetry.

Off-shell $k_f \neq k_i$ and $ \sigma_1 \cdot \sigma_2 $ enters as an
independent additional operator combination, since the on-shell
identity, Eq.~(\ref{eq:on-id}) does not hold. Another feature is that
the off-shell independent components $V_P$ and $W_P$ are significantly
non-vanishing.

\subsection{The separation distance}

Once the form of the (semi-local) potential is specified, the radial
dependence of the components $V_n(r)$ must be determined. We have
taken a {\it sharp} separation distance, $r_c$, where we distinguish
between the known and calculable part of the potential, by invoking
QFT with hadronic degrees of freedom, and the unknown part which
should not be calculable at the hadronic level, since it involves
finite size, quark overlap and exchange terms, etc.  Obviously, for a
model independent analysis this separation distance should not be
smaller than the {\it elementary} radius, $r_e$, i.e., the distance
above which particles may be regarded as point like (see e.g.
Fig.~\ref{Fig:VU-r}). The analysis of Ref.~\cite{Arriola:2016hfi}
based on quark cluster considerations and the nucleon and $N \Delta$
transition form factors suggest that also that $r_e \sim 1.8$fm. Thus,
the separation reads,
\begin{equation}
V(r) = V_{\rm short} (r) \theta (r_c - r ) + V_{\rm QFT} (r) \theta
(r- r_c) \, ,
\label{eq:sep}
\end{equation}
Although it has become customary to use smooth radial functions, their
particular shape may mask finite size features, and thus in all our
studies we insist on this sharp separation scheme\footnote{The
  essential issue here is not the smoothness of the separation
  function (in our case a step function), but rather the distance
  where the full QFT deduced potential is switched on. A further
  advantage of this sharp separation is the applicability of
  two-potential formulas~\cite{PavonValderrama:2009nn}. Moreover,
  analytical properties depend on $V_{\rm QFT} (r)$
  only~\cite{RuizdeElvira:2018hsv}.}. While the long range part
contains and discriminates between strong and electroweak
contributions, the short range part contains {\it both} contributions
and we will make no effort to disentangle them explicitly.

\subsection{The number of independent parameters}

In order to deal with the short distance components we may in principle
propose some functions with ``reasonable'' shapes characterized by some
parameters as it is, e.g., the case of the AV18
potential~\cite{Wiringa:1994wb}. As we will discuss below, uncertainties in
the potential are actually dominated by the arbitrariness in the
representation of the potential at short distances. To overcome this
difficulty we have appealed in our analyses to the use of coarse grained
potentials, a scheme proposed long ago by Avil\'es~\cite{Aviles:1973ee} and
rediscovered in Ref.~\cite{Entem:2007jg} (see
e.g.Ref.~\cite{NavarroPerez:2011fm,RuizArriola:2019pwt} for  pedagogical
discussions).  This approach is inspired by a Wilsonian point of view and an
optimal sampling of the potential is implemented after Nyquist theorem. We
take a grid of equidistant radial ``thick'' points in coordinate space
separated by the finite resolution given by the shortest de Broglie
wavelength, $\Delta r = \hbar /p_{\rm CM}^{\rm max} \sim 0.6 {\rm fm}$ up to
the radius $r_c= 3 \, {\rm fm}$, above which charge dependent $1\pi$ exchange
gives the entire strong contribution. This gives, for instance, for an S-wave
$r_c/\Delta r=5$ thick points. A simple calculation including all active
partial waves, $ L \lesssim p r_c $, yields the
estimate~\cite{Fernandez-Soler:2017kfu,RuizArriola:2019pwt},
\begin{equation}
  N_{\rm Par} \sim \frac12 (p_{\rm CM}^{\rm max} r_c)^2 \, g_S \, g_T \, , 
  \label{eq:npar}
\end{equation}
where $g_S$ and $g_T$ are spin and isospin degeneracy factors. The counting of
parameters for pp and np~\cite{Perez:2013cza} yields about 40 ``thick'' points
$r_n$ if the fit is carried up to a maximal $T_{\rm LAB} \le 350$MeV. This
{\it a priori} estimate coincides with the bulk of parameters which have been
needed to fit data satisfactorily in the past.  The simplest way these thick
points may be represented by delta-shells (DS)~\cite{NavarroPerez:2012qf} as
originally proposed by Avil\'es~\cite{Aviles:1973ee}, and the potential values
at these points $V_i (r_n)$ are taken as the fitting parameters, 
\begin{equation}
  V_{\rm short} (r) = \sum_{n,i} \Delta r \, O_n \, V_n (r_i) \delta
  (r -r_i) \, .
\end{equation}
Of course, the estimate Eq.~(\ref{eq:npar}), shows that in order to
minimize the number of parameters for a given maximal momentum the
value of $r_c$ should be taken as smallest as possible, but never
smaller then the elementary radius, $r_e \sim1.8$fm.

\subsection{The long range contributions}

The hadronic QFT calculable contribution is separated into two pieces, the
strong (pion exchange) piece and the purely EM piece, 
\begin{equation}
  V_{\rm QFT} =  V_{\pi}(r) + V_{\rm EM}(r) \, . 
\end{equation}
The charge dependent OPE potential in the long range part of the interaction
is the same as the one used by the Nijmegen group on their 1993 PWA~\cite{Stoks:1993tb} and reads
\begin{equation}
 V_{m, \rm OPE}(r) = f^2\left(\frac{m}{m_{\pi^\pm}}\right)^2\frac{1}{3}m\left[Y_m(r){\mathbf \sigma}_1\cdot\mathbf{\sigma}_2 + T_m(r)S_{1,2} \right]
\end{equation}
being $f$ the pion coupling constant, ${\mathbf \sigma}_1$ and ${\mathbf
\sigma}_2$ the single nucleon Pauli matrices, $S_{1,2}$ the tensor operator,
$Y_m(r)$ and $T_m(r)$ the usual Yukawa and tensor functions,
\begin{align}
 Y_m(r) &= \frac{e^{-m r}}{m r}, \nonumber \\
 T_m(r) &= \left(1+ \frac{3}{mr} + \frac{3}{(mr)^2} \right)\frac{e^{-m r}}{m r}. 
 \label{eq:Yukawa}
\end{align}
Charge dependence is introduced by the difference between the charged
$m_{\pi^\pm}$ and neutral $m_{\pi^0}$ pion mass by setting
\begin{align}
 V_{{\rm OPE},pp}(r) &= V_{m_{\pi^0},\rm OPE}(r), \nonumber \\
 V_{{\rm OPE},np}(r) &= -V_{m_{\pi^0},\rm OPE}(r)+ (-)^{(T+1)}2V_{m_{\pi^\pm},\rm OPE}(r).
 \label{eq:BreakIsospinOPE}
\end{align}

The neutron-proton electromagnetic potential includes only a magnetic moment
interaction
\begin{equation}
V_{\rm EM, np}(r) = V_{\rm MM, np}(r) = -\frac{\alpha \mu_n}{2M_{n} r^3}  \left( \frac{\mu_{p}S_{1,2}}{2 M_p}  + \frac{{\bf L}\!\cdot\!{\bf S}}{\mu_{np}}  \right),
\end{equation}
where $\mu_n$ and $\mu_p$ are the neutron and proton magnetic moments, $M_n$
the neutron mass, $M_p$ the proton one and ${\bf L}\!\cdot\!{\bf S}$ is the
spin orbit operator. The EM terms in the proton-proton channel include one and
two photon exchange, vacuum polarization and magnetic moment,
\begin{equation}
 V_{\rm EM, pp}(r) = V_{\rm C1}(r) + V_{\rm C2}(r) + V_{\rm VP}(r) + V_{\rm MM, pp}(r)
\end{equation}
where
\begin{align}
  &V_{\rm C1}(r) = \frac{\alpha'}{r} \ ,     \\
  &V_{\rm C2}(r) = -\frac{\alpha\alpha'}{M_{p} r^2} \ ,  \\
  &V_{\rm VP}(r) = \frac{2\alpha\alpha'}{3\pi r} \int^{\infty}_{1} e^{-2m_{e}rx}\left(1+\frac{1}{2x^{2}}\right)
            \frac{\sqrt{x^{2}-1}}{x^{2}}dx  \ ,                 \\
%            \frac{(x^{2}-1)^{1/2}}{x^{2}} \ ,                 \\
  &V_{\rm MM, pp}(r) = -\frac{\alpha}{4M^{2}_{p} r^3} \left[
                 \mu^{2}_{p}S_{1,2}  + 2(4\mu_{p}-1){\bf L}\!\cdot\!{\bf S}  \right].
\end{align}  
Note that these potentials are {\it only} used above $r_c = 3 {\rm fm}$ and
thus form factors accounting for the finite size of the nucleon can be set to
one.  Energy dependence is present through the parameter
\begin{equation}
  \alpha'= \alpha\frac{1+2k^2/M_p^2}{\sqrt{1+k^2/M_p^2}} \, , 
\end{equation}
where $k$ is the center of mass momentum and $\alpha$ the fine structure
constant. 

\section{The partial wave analysis}

The great achievement of the Nijmegen group 25 years ago was to
provide for the {\it first time} a statistically satisfactory
description of a large amount of scattering
data~\cite{Stoks:1993tb,Stoks:1994wp}. This was possible because of
two good reasons. First, charge dependence (CD) and tiny
electromagnetic effects such as vacuum polarization, magnetic moments
interactions (which requires summing up over thousend partial waves)
and relativistic corrections to the Coulomb scattering were
incorporated. Second, a suitable selection of all the data was
implemented.  The Granada-2013 database is based on a similar
approach, but with two significant improvements: the number of data is
almost twice and the selection process has been made self-consistent
as suggested by Gross and Stadler~\cite{Gross:2008ps}. A summary of
the situation is illustrated by the abundance plots in
Fig.~\ref{fig:NN-abundance}.

\subsection{Validation and Falsification: Frequentist vs Bayesian}

In low energy nuclear physics a great deal of significant information
is extracted by analyzing data (see
e.g.~\cite{Dobaczewski:2014jga}). Thus, making first a fair
statistical treatment of NN is an absolute precondition to aim at any
subsequent precision goal in {\it ab initio} calculations. This
applies in particular to chiral interactions and their
validation/falsification, as we will discuss below. We remind the fact
that least squares $\chi^2$-fitting any (good or bad) model to some
set of data is always possible and corresponds to just minimizing a
distance between the predictions of the theory and the experimental
measurements. The crucial aspect is the {\it statistical} significance
of the fit. If we take $\chi^2$ as a function of the short distance
potential components at the grid points $V_i(r_n)$,
\begin{equation}
\chi_{\rm min}^2= \min_{V_i(r_n)} \chi^2 (V_i (r_n))= \sum_{i=1}^{N_{\rm Dat}} \left[ \frac{{\cal O}_i^{\rm th} (V_i (r_n)  ) -{\cal O}_i^{\rm exp}}{\Delta {\cal O}_i^{\rm exp}} \right]^2 
\end{equation}
The essential point is whether or not the discrepancies between theory and
experiment are statistical fluctuations which might be improved by making
better measurements. Namely, we ask if the residuals
\begin{equation}
R_i = \frac{{\cal O}_i^{\rm th}|_{\rm min} - {\cal O}_i^{\exp}}{\Delta {\cal O}_i^{\rm exp}}   \qquad , \, i=1, \dots, N_{\rm Dat} 
\end{equation}
follow a normal distribution. Of course, we can never be certain about this,
but the statistical approach provides one probabilistic answer and depends on
i) the number of data, $N_{\rm Dat}$, ii) the number of parameters determined
from this data, $N_{\rm Par}$, and iii) the nature of experimental
uncertainties~\cite{Perez:2014yla,Perez:2015pea}.

In the Bayesian approach one poses the natural question: What is the
probability $P(T/D)$ that given the data $D$ the theory $T$ occurs?.
This requires some {\it a priori} probabilistic expectations, $P(T)$,
on the goodness of the theory regardless of the data and is dealt with
often by augmenting the experimental $\chi_{\rm exp}^2$ with an
additive theoretical contribution $\chi_{\rm th}^2$.  However, it can
be proven that when $N_{\rm Dat} \gg N_{\rm Par}$ one can ignore these
a priori expectations since $ \chi^2_{\rm exp} \sim N_{\rm Dat} \gg
\chi_{\rm th}^2 \sim N_{\rm Par} $ and proceed with the Frequentist
approach where just the opposite question is posed: what is the
probability $P(D/T) $ of measuring data $D$ given the theory
$T$?. Bayes theorem stating that the joint probability $P(D, T)=P(D/T)
P(T)= P(T/D) P(D)$ provides the coneection. One could stay Bayesian if
some relative weighting of $\chi_{\rm exp}^2$ and $\chi_{\rm th}^2$ is
implemented (see~\cite{Ledwig:2014cla,Perez:2016vzj} and references
therein). In our analysis below, where we have $N_{\rm Dat} \sim 8000$
( see Fig.~\ref{fig:NN-abundance} ) and $N_{\rm Par} \sim 40$ (see
Eq.~(\ref{eq:npar}), we expect no fundamental differences between the
Bayesian and frequentist approaches. Note that 1) we can never be sure
that the model is true and 2) any experiment can be right if errors
are sufficiently large and in this case the theory cannot be
falsified.
% This said, $p=0.68$ when $\chi^2/\nu = 1 \pm \sqrt{2/\nu}$ with $\nu=N_{\rm
% Dat}-N_{\rm Par}$.

In general we expect discrepancies between theory and data and,
ideally, if our theory is an approximation to the true theory we
expect the optimal accuracy of the truncation to be comparable with
the given experimental accuracy and both to be compatible within their
corresponding uncertainties (see \cite{Wesolowski:2015fqa} for a
Bayesian viewpoint).  If this is or is not the case we validate or
falsify the approximated theory against experiment and declare theory
and experiment to be compatible or incompatible respectively. Optimal
accuracy, while desirable, is not really needed to validate the
theory. In the end largest errors dominate regardless of their origin
(or their Bayesian justification); the approximated theory may be
valid but inaccurate.  We will see below that this is the case for
currently existing chiral interactions, where the truncation error is
{\it at best} comparable to the spread of statistically verified
interactions against the NN database.

\begin{figure*}[htb]
\begin{center}
\includegraphics[width=\linewidth]{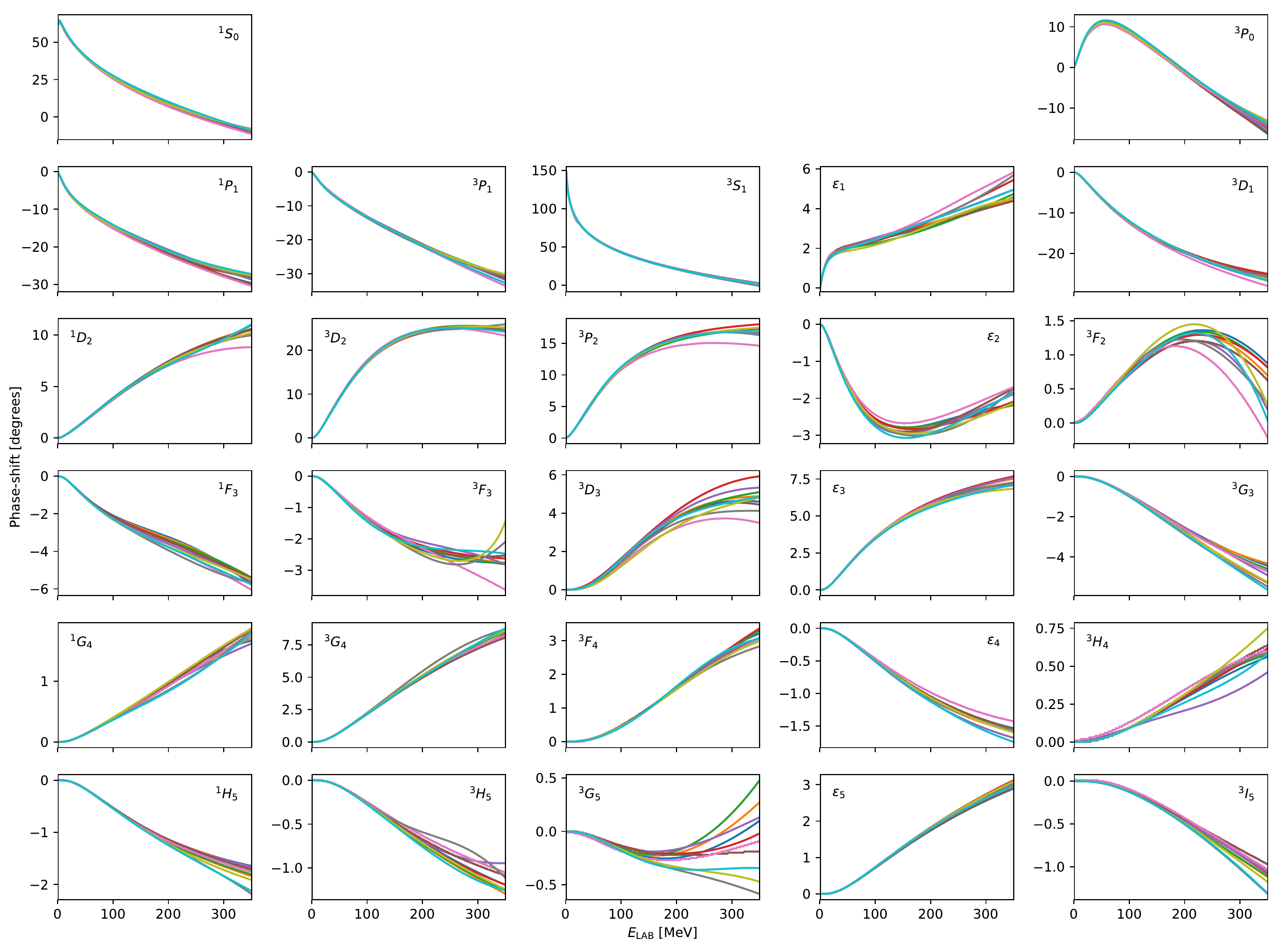}
\end{center}
\caption{\label{fig:ps-sys} np Phase shifts for 10 high quality fits in all
partial waves with $J \le 5$ as a function of the LAB energy. We show the
Nijmegen PWA~ Nijm I, NijmII Reid93~\cite{Stoks:1994wp} the
AV18~\cite{Wiringa:1994wb}, CD Bonn~\cite{Machleidt:2000ge},
Spectator~\cite{Gross:2008ps}, and the Granada
DS-OPE~\cite{Perez:2013mwa,Perez:2013jpa},
DS-$\chi$TPE~\cite{Perez:2013oba,Perez:2013cza} and
Gauss-OPE~\cite{Perez:2014yla}.}
\end{figure*}

\subsection{Fitting and selecting data}

This approach allows to select the largest self-consistent existing NN
database with a total of 6713 NN scattering data driven by the coarse grained
potential~\cite{Perez:2013jpa,Perez:2014yla} with the rewarding consequence
that statistical uncertainties can confidently be propagated~\footnote{The
resulting Granada-2013 can be downloaded from \\ (\url{http://www.ugr.es/~amaro/nndatabase/}).}. Precise determinations of chiral
coefficients, $c_1,c_3,c_4$~\cite{Perez:2014bua,Perez:2013za}, the isospin
breaking pion-nucleon~\cite{Perez:2016aol,Arriola:2016hfi}, and the
pion-nucleon-delta~\cite{Perez:2014waa} coupling constants have been made.

One important aspect regards the correlation properties among the
fitting parameters. In our case one gets for the potential components
at the sampling points $r_n$ the values $V_i (r_n) \pm \Delta V_i
(r_n)$. When going to the partial wave basis, $V_{l,l'}^{JS} (r_n)$
one observes that different partial waves are largely
decorrelated~\cite{Perez:2014yla,Perez:2014kpa}. This shows {\it a
  posteriori} that the assumption of independent parameters is
somewhat confirmed. As a bonus, this lack of strong correlations
allows in practice for a very efficient search of the minimum of the
$\chi^2$.

\subsection{Systematic vs Statistical errors}

Our analysis has clearly demonstrated that at least the statistically
validated 6-Granada potentials exhibit {\it similar} statistical
errors but {\it different} most likely values of the predicted
quantities~\cite{Perez:2014waa}. The most vivid example is provided by
the phase shifts as we show in Fig.~\ref{fig:ps-sys}. We also plot in
the figure the results for the previous high quality potentials at
their time (defined by their $\chi^2/N \sim 1$). Our observation is
that for the Granada potentials, which are statistically validated
with the {\it same} Granada-2013 database, we have that if a
phase-shift for potential $V^{(i)}$ in a given partial wave is
$\delta^{(i)} \pm \Delta \delta_{\rm stat}^{(i)} $, then
\begin{equation}
 \Delta \delta_{\rm stat}^{(1)} \sim \dots \sim  \Delta \delta_{\rm stat}^{(6)} \, , 
\end{equation}
but the standard deviation defined as usual, obbeys  
\begin{equation}
\Delta \delta_{\rm sys} \equiv {\rm Std} (\delta^{(1)}, \dots ,
  \delta^{(6)}) \gg \Delta \delta_{\rm stat}^{(i)} \, .
\end{equation}
In all cases the potential {\it above} $r=3$fm including CD-OPE and all
electromagnetic effects are the {\it same}, thus the discrepancies between
different forms of representing the potential at short distances dominate the
uncertainties, rather than the np and pp experimental data themselves. 

A way of comparing potentials which actually minimizes the role of short
distances is by considering radial moments,
\begin{equation}
C_{n,i} \sim \int_0^\infty d^3 r r^n V_i(r)
\end{equation}
which correspond to a low momentum expansion of the potential in
momentum space. The resulting values for all active channels have been
found to be rather universal yet the larger systematic vs statistical
uncertainties are also found~\cite{Perez:2016vzj}. This effect is also
found in the low energy parameters characterizing low energy
scattering, see Eq.~(\ref{eq:ERE-coup}), and the Wolfenstein
parameters~\cite{RuizSimo:2017anp}.

\section{The chiral potentials}

In the previous sections we have reviewed some aspects needed in the
construction of the Granada-2013 database. In the remaining sections we will
compare the quality of the chiral approach on the light of these
developments. We will try to check to what extent the chiral hierarchy
complies with the trends observed in the NN data analysis in the last 26 years
since the Nijmegen group first found a statistically satisfactory description
of the data at the time with a large database.

\subsection{Chiral counting}

Despite the long discussions on the correctness and suitability of the
Weinberg power counting, most applications to light nuclei use this
scheme based on a NN potential in practice, and we will therefore
restrict our short reminder to this very significant case.  The
Weinberg chiral counting relies on the heavy-baryon formalism and a
naive dimensional analysis of the NN potential.  Thus, according to
the Feynman rules of covariant perturbation theory, a nucleon
propagator is $Q^{-1}$, a pion propagator $Q^{-2}$, each derivative in
any interaction is $Q$, each four-momentum integration $Q^4$ and the
pion mass $m_\pi$ is also $Q$. In addition, for reasons explained in
Ref.~\cite{Weinberg:1990rz}, terms including factors of $Q/M_N$, where
$M_N$ denotes the nucleon mass are counted by the rule $Q/M_N \sim
(Q/\Lambda_\chi)^2$. There are two kinds of contributions for the
total potential
\begin{equation}
V= V_{\rm ct} + V_\pi 
\end{equation}
where $V_\pi $ is the long range contributions containing an increasing number
of pion exchanges,
\begin{equation}
V_{\pi} = V_{1\pi} + V_{2\pi} + V_{3\pi} + \ldots \,,
\end{equation}
which are discontinuous function of the momentum transfer at $q = i m_\pi, 2 i
m_\pi, 3 i m_\pi, \dots $ and $V_{\rm ct}$ are contact contributions which do
not present this discontinuity and to a given order reduce to polynomials in
initial and final momenta, $p$ and $p'$ respectively, 
\begin{equation}
V_{\rm ct} =
V_{\rm ct}^{(0)} + 
V_{\rm ct}^{(2)} + 
V_{\rm ct}^{(4)} + 
V_{\rm ct}^{(6)} 
+ \ldots \; ,
\label{eq_ct}
\end{equation}
where the superscript denotes the power or order and 
parity and time-reversal only allows even powers of momentum.
Pion exchange terms, we have a similar expansion 
\begin{align}
V_{1\pi} & =  V_{1\pi}^{(0)} + V_{1\pi}^{(2)} 
+ V_{1\pi}^{(3)} + V_{1\pi}^{(4)} + V_{1\pi}^{(5)} + \ldots 
\label{eq_1pe_orders}
\\
V_{2\pi} & =  V_{2\pi}^{(2)} + V_{2\pi}^{(3)} + V_{2\pi}^{(4)} + V_{2\pi}^{(5)} 
+ \ldots \\
V_{3\pi} & =  V_{3\pi}^{(4)} + V_{3\pi}^{(5)} + \ldots \,,
\end{align}
Order by order, the complete $NN$ potential builds up as follows:
\begin{align}
V_{\rm LO} & = V_{\rm ct}^{(0)} + 
V_{1\pi}^{(0)} 
\label{eq_VLO}
\\
V_{\rm NLO} & = V_{\rm LO} +
V_{1\pi}^{(2)} +
V_{2\pi}^{(2)}+V_{\rm ct}^{(2)} 
\label{eq_VNLO}
\\
V_{\rm N2LO} & = V_{\rm NLO} +
V_{1\pi}^{(3)} + 
V_{2\pi}^{(3)} 
\label{eq_VNNLO}
\\
V_{\rm N3LO} & = V_{\rm N2LO} +
V_{1\pi}^{(4)} + 
V_{2\pi}^{(4)} +
V_{3\pi}^{(4)} +V_{\rm ct}^{(4)} 
\label{eq_VN3LO}
\\
V_{\rm N4LO} & = V_{\rm N3LO} +
V_{1\pi}^{(5)} + 
V_{2\pi}^{(5)} +
V_{3\pi}^{(5)} 
\label{eq_VN4LO}
\end{align}
where LO stands for leading order, NLO for next-to-leading order, etc.. From
the point of view of the potential operators, Eq.~(\ref{eq_nnamp}),  the chiral hierarchy reads 
\begin{align}
  m_\pi^2 W_T &= {\cal O} (Q^0) \\ V_T, W_C, V_S &= {\cal O} (Q^2) \\ V_C,W_S, 
  m_\pi^2 V_{SL}, m_\pi^2 W_{SL} &= {\cal O} (Q^3) \\
  m_\pi^4 V_Q &= {\cal O} (Q^5) \\
  m_\pi^4 W_Q, m_\pi^2 V_P, m_\pi^2 W_P & \le  {\cal O} (Q^6) 
\end{align}
where we have multiplied by a power of $m_\pi$ so that all components
have the same dimensions.

It is noteworthy that the central force, traditionally taken as the main
contribution to the mid range attraction which holds atomic nuclei together,
rather than being LO becomes N2LO in the chiral expansion\footnote{ This has
lead to explore other QCD based approaches to the nuclear force such as e.g.
the large $N_c$ expansion, where $V_C$ (together with $W_T$ ) becomes leading
order and the remaining components are $1/N_c$ suppressed by a relative
$1/N_C^2$~\cite{Kaplan:1995yg,Kaplan:1996rk,CalleCordon:2008cz}. It would be
interesting to check if the chiral hierarchy is followed by phenomenological
potentials.}. Moreover, we would need to go at least to order ${\cal O} (Q^7)$
or higher, to generate a term $W_Q$. We will see later on that the
phenomenological approaches require a non-vanishing and statistically
significant $|W_Q | \gg \Delta W_Q$.

The size of the potential is relevant to discuss the chiral hierarchy, but it
is unclear how this could be done quantitatively. One may, for instance,
compare local potentials at a fixed distance, but the question ambiguity
remains as the question of what is a reasonable distance to compare?. Field
theoretical potentials diverge at short distances and they even make the
quantum mechanical problem mathematically ill defined. The separation distance
between the explicit pion exchange and the unknown part of the nuclear
potential plays the role of a cut-off in coordinate space. We believe that the
comparison of the local components of potentials as a function of the momentum
transfer made in Fig.~\ref{fig:mom-sys-local} goes more to the point. These
components can be related to the contributions displayed in Eq.~(\ref{eq_nnamp})
by the relations in the Appendix~\ref{sec:mom-sp}. In fairness the comparison
should be made for small $q$ values (typically $\sim m_\pi$).  The central
component makes a large variation from this range where it is small to a large
value for $q \sim 3 {\rm fm}^{-1} \sim 4 m_\pi$. This is in agreement with the
chiral suppression {\it and} the phenomenology on the dominance of the
mid-range central force. At small q the largest component is by far the tensor
force as implied also by the chiral counting.
  % $W_{Q}$ ...

\subsection{Scale dependence and the number of parameters}

The short range components of the NN potentials {\it depend} on the
maximal fitting energy (see e.g. Ref.~\cite{NavarroPerez:2013iwa} for
an explicit example). This is not so often realized because the
maximal fitting energy is usually fixed to a conventional value,
usually $T_{\rm LAB} \le 350$MeV. If we take $\Lambda=p_{\rm CM}^{\rm max}$, the
LS equation at the partial waves level reads,
\begin{eqnarray}
  T_{l'l}^{JS}(p',p) &=& V_{l'l}^{JS}(p',p) \nonumber \\ &+& \frac{2}{\pi} \sum_{l''}\int_0^\Lambda dq V_{ll"}^{JS} (p',q) \frac{q^2}{k^2-q^2} T_{l'',l}^{JS}(q,p) \,  
\end{eqnarray}
In a loose sense, the contact pieces in the chiral potentials
represent the role of counter-terms in quantum field theory, but their
significant value depends on $\Lambda$. For small momenta one has
\begin{equation}
  V_{l',l}^{JS} (p',p) |_{ct} = (p')^{l'} p^l \sum_{n,m=0}^\infty C_{nm}(\Lambda)  (p')^{2n} p^{2m}
  \label{eq:vlowk}
\end{equation}
where $C_{nm}=C_{mn}$ are real. In the limit, $\Lambda \to 0$, their
values may be directly computed from the low energy threshold
parameters~\cite{RuizArriola:2016vap} only up to ${\cal O} (p^2)$. For
higher orders, there are more counter-terms than threshold parameters,
featuring again in the low energy limit the ambiguity of the inverse
scattering problem. Some ambiguity can be shifted by performing a
suitable unitary transformation as, for instance, discard purely
off-shell contributions~\cite{Reinert:2017usi} in which case to ${\cal
  O} (p^{2n})$ the number of parameters would be $N_{\rm
  Par}=(4+5n)(n+1)/2=2,9,21,38,60,\dots$ for $n=0,1,2,3,4,
\dots$~\cite{PavonValderrama:2005ku}.

\subsection{Chiral potential fits to NN scattering data}

While the most portable form of a PWA corresponds to listing the phase shifts
and this clearly provides a very reasonable picture of the scattering process,
from the point of view of validation/falsification of the interaction fitting
phase-shifts and fitting scattering data are quite different. Therefore we will
only consider chiral fits to a full database containing cross sections and
polarization observables. The first chiral fits using the N2LO chiral
potentials~\cite{Kaiser:1997mw} to the Nijmegen database were undertaken by
the Nijmegen group itself for pp~\cite{Rentmeester:1999vw} and for
pp+np~\cite{Rentmeester:2003mf}, allowing for a determination of the chiral
constants $c_1,c_3,c_4$. The newest generation of chiral potentials provide
fits to the Granada-2013
database~\cite{Perez:2013oba,Perez:2013cza,Perez:2014bua,Carlsson:2015vda,Piarulli:2016vel,Reinert:2017usi,Entem:2017gor}.
The best fit we have obtained provides $\chi^2/\nu=1.025$ when isospin
breaking is incorporated both in the pion masses, the coupling constants and
in the short range contribution~\cite{Perez:2016aol,Arriola:2016hfi}.

The recent calculations by the Idaho-Salamanca group
\cite{Entem:2017gor} and the Bochum group~\cite{Reinert:2017usi}, are
possibly the most complete analyses to date of chiral potentials going
up to N4LO in the Weinberg counting which basically use the Granada
2013 database which contains $N_{\rm Dat}=2996_{pp}+ 3717_{np}= 6713$
data which are $3\sigma$ mutually compatible by a coarse grained
potential with $\chi^2 / \nu = 1.027$. The fact that both groups take
$T_{\rm LAB} \lesssim 290-300$MeV and that they thus have about
$N_{\rm Dat} \sim 4850$ implies that a fit would be satisfactory {\it
  a posteriori} if $\chi^2/N_{\rm Dat}=1 \pm 0.02$. Our discussion
below adopts this point of view.

One remarkable aspect of the Bochum analysis~\cite{Reinert:2017usi} is
the fact that they show that for the first time the chiral potential
fit outperforms the traditional potentials such as those of the
Nijmegen group and AV18 with much less parameters when the
Granada-2013 database is taken. In both Idaho-Salamanca and Bochum
cases cases a relevant (Gaussian) regulator dependence in the quality
of the fit in the ``reasonable'' range $\Lambda \sim 450-550$MeV is
reported. Likewise, a systematic order by order analysis seems to
indicate convergence of the chiral approach.

In the Idaho-Salamanca study a N4LO nonlocal potential is used to fit $N_{\rm
Dat}=4853$-NN data are analyzed below $T_{\rm LAB}=290$MeV. The quality of
such a fit is not given, but they quote the results for the N4LO$^+$ consisting
in the addition of additional counter-terms in the F-waves, which are nominally
N5LO, but disregarding the presumably small pion contributions $V_\pi^{(6)}$,
\begin{equation}
V_{{\rm N4LO}^+} \equiv V_{\rm N4LO} + V_{\rm ct}^{(6)}  
\end{equation}
obtaining in this case a total best fit value $\chi^2/N_{\rm Dat}=1.15$ when
148 np data (Granada-accepted by the $3\sigma$ criterion) are replaced by 140
pp data (Granada-rejected by the $3\sigma$ criterion).  Likewise,
the Bochum group uses a semi-local N4LO chiral potential and analyze a total of
$N_{\rm Dat}=4855$-NN data are analyzed below $T_{\rm LAB}=300$MeV with a best
$\chi^2/N_{\rm Dat}=1.39$. Only after some (Granada accepted) data are removed
and adding also F-waves counter-terms, thus working in the N4LO$^+$ scheme, they
get $\chi^2 /N_{\rm Dat} =1.03 $ which is very good. From this we may infer
that the effect of going from N4LO to (incomplete) N4LO$^+$ is crucial for
claiming that the chiral expansion has converged.

We believe that this selected but different exclusion best exemplifies
the main difference between power counting and the coarse graining
approach used to fix the Gra\-na\-da\--\-2013 database\footnote{An
  argument has been made (see also \cite{Ekstrom:2013kea}) to exclude
  some old but high accuracy cross section data from the Granada-2013
  database to the highest considered order, which is assumed to be
  less accurate than the data.}. Of course, it would be highly
desirable to go to the next order and check if the problem could be
solved. An approximate way of doing this would be to include the
G-waves counter-terms in a N4LO$^{++}$ scheme.

To summarize, both calculations are nominally N4LO but the quality of
the fits only gets improved when F-wave counter-terms are added when
$T_{\rm LAB} \lesssim 290-300$MeV, the so-called N4LO$^+$, which
indicate a flagrant violation of the standard Weinberg power counting
unless it is shown the $V_\pi^{(6)}$ is significantly
negligible. Another possibility is to reduce the energy, so that
F-wave counter-terms become marginal. We comment this possibility
below in Sect.~\ref{sec:sdc}.

As it has been emphasized over the years by their own practitioners
the {\it raison d'\^etre} of the EFT approach lies in the systematic
determination of the physical observables in a power counting
scheme. Taken at face value, the quoted values of the $\chi^2/\nu$ are
several $\sigma$'s away from their expected values, showing that the
chiral N4LO$^+$ potentials despite violating the chiral counting, are
still incompatible with the existing data. The quality of the fits in
the N4LO$^+$ case, supports their use in nuclear physics as valid
representations of the NN scattering data below $T_{\rm
  LAB}=290-300$MeV, but not more than say any of the previous high
quality potentials since the 1993 Nijmegen analysis.

In our works~\cite{Perez:2014yla,Perez:2014kpa} (see also \cite{Perez:2015pea}
for a $\pi\pi$ based presentation), we have insisted on the fact that if
$\chi^2/ \nu$ is not within the expected values, one can still slightly scale
the errors by the so-called Birge factor {\it provided} one can show that the
residuals are still a scaled Gaussian distribution with a variety of
statistical normality tests based in an analysis of the tails or the moments
of the distribution of residuals. For instance, for $\nu=7000$ one should have
$\chi^2/\nu = 1 \pm 0.017$.  Having instead, $\chi^2/\nu = 1.08$ would mean
$5\sigma$ incompatible fit. However, if the normality test is passed it is
plausible that the rescaling of the $\chi^2$ corresponds to a rescaling of the
data uncertainties by only 10$\%$ factor, fully compatible with the ``error on
the error''. We of course, should not exclude the interesting possibility that
the N4LO analyses themselves pass the normality tests and thus qualify as
suitable description of NN scattering data.

\begin{table*}
\caption{\label{tab:fpf0fc-TPE} The pion-nucleon coupling constant $f^2 = f_p^
2=f_0^ 2=f_c^ 2$ and the chiral constants $c_1$, $c_3$ and $c_4$ determined
from different fits to the Granada-2013 database and of the CD-OPE plus $\chi
$TPE potential depending on the cut-off radius $r_c$.  Charge dependence is
only allowed on the $^1S_0$ partial wave. Here we take $T_{\rm LAB} \le 350
{\rm MeV}$ and define $N_\sigma = (\chi^2/\nu -1)/\sqrt{2/\nu}$.} % For LaTeX tables use

\begin{tabular*}{\textwidth}{@{\extracolsep{\fill}} c D{.}{.}{1.6} D{.}{.}{4.7} D{.}{.}{4.7} D{.}{.}{2.6} ccccccc D{.}{.}{2.3}}
    \hline\noalign{\smallskip}
    $r_c ({\rm fm})$& 
    \multicolumn{1}{c}{$f^2$} & 
    \multicolumn{1}{c}{$c_1 ( {\rm GeV}^{-1})$} & 
    \multicolumn{1}{c}{$c_3 ( {\rm GeV}^{-1})$} & 
    \multicolumn{1}{c}{$c_4 ( {\rm GeV}^{-1})$} & 
    $\chi^2_{pp}$ &
    $\chi^2_{np}$ &
    $\chi^2$ &
    $N_{\rm Dat}$ &
    $N_{\rm Par}$ &
    $\chi^2/\nu $ &
    \multicolumn{1}{c}{$ N_\sigma $} \\ 
    \noalign{\smallskip}\hline\noalign{\smallskip}
    %\colrule
    % \hline 
    3.6 & 0.075      & 1010.0(306) & -990.9(264) &   9.6(140) & 2975.09 & 3879.15 & 6854.24 & 6719 & 63 & 1.030 &  1.7 \\ 
    3.6 & 0.0710(6)  &  978.3(390) & -961.1(353) &  -4.0(148) & 2965.28 & 3869.62 & 6834.90 & 6719 & 64 & 1.027 &  1.6 \\
    %% 3.6 & 0.0694(6)  &   -0.76     &  -29.2(27)  & -24.4(150) & 3034.81 & 3881.20 & 6916.01 & 6719 & 63 & 1.039 &  2.3 \\ 
    %\hline  
    3.0 & 0.075      &  -44.4(70)  &   39.5(51)  &  -4.4(26)  & 2979.46 & 3980.27 & 6959.73 & 6721 & 49 & 1.043 &  2.5 \\ 
    3.0 & 0.0763(3)  &  -35.2(79)  &   31.3(60)  &  -6.4(27)  & 2983.95 & 3968.28 & 6952.23 & 6721 & 50 & 1.042 &  2.4 \\ 
    %% 3.0 & 0.0765(2)  &   -0.76     &    3.4(4)   &  -8.1(26)  & 2987.37 & 4004.52 & 6991.89 & 6722 & 49 & 1.048 &  2.8 \\
    %\hline 
    2.4 & 0.075      &  -10.6(18)  &    5.2(10)  &  -2.1(8)   & 3064.38 & 4049.88 & 7114.26 & 6718 & 41 & 1.065 &  3.8 \\ 
    2.4 & 0.0748(2)  &  -11.9(20)  &    6.0(12)  &  -2.3(9)   & 3065.80 & 4048.30 & 7114.11 & 6718 & 42 & 1.066 &  3.8 \\ 
    %% 2.4 & 0.0749(1)  &   -0.76     &   -1.5(1)   &  -1.9(8)   & 3069.04 & 4058.94 & 7127.98 & 6719 & 41 & 1.067 &  3.9 \\
    %\hline 
    1.8 & 0.075      &   -1.9(6)   &   -3.7(2)   &   4.4(2)   & 3101.24 & 4059.32 & 7160.56 & 6717 & 33 & 1.071 &  4.1 \\ 
    1.8 & 0.0763(2)  &   -1.6(6)   &   -3.7(3)   &   4.3(2)   & 3077.00 & 4050.22 & 7127.22 & 6717 & 34 & 1.066 &  3.8 \\
    %% 1.8 & 0.0764(1)  &   -0.76     &   -4.25(5)  &   4.3(2)   & 3085.99 & 4055.16 & 7141.15 & 6718 & 33 & 1.068 &  3.9 \\
    %\hline 
    1.2 & 0.075      &  -11.17(9)  &    0.76(2)  &   2.822(2) & 3428.38 & 4659.52 & 8087.90 & 6715 & 25 & 1.209 & 12.1 \\
    1.2 & 0.07500(3) &  -11.17(9)  &    0.76(3)  &   2.821(6) & 3428.28 & 4659.02 & 8087.31 & 6715 & 26 & 1.209 & 12.1 \\
    %% 1.2 & 0.07385(3) &   -0.76     &   -3.592(4) &   3.25(5)  & 3769.83 & 4732.91 & 8502.74 & 6713 & 25 & 1.271 & 15.7 \\
    \noalign{\smallskip}\hline
    %\botrule
  \end{tabular*}
\end{table*}

\subsection{N5LO chiral forces}

The chiral expansion provides a hierarchy in the components of the NN
interaction. Can we check this hierarchy by just analyzing NN scattering data
{\it directly}?. We feel that this is rather difficult, however, we can
instead check whether the chiral pattern is verified for potential models.  In
the Weinberg counting to sixth order (N5LO ) one finds that the component
$W_Q=0$ whereas one has $V_Q={\cal O} (1/f^4 M_N^2)$~\cite{Kaiser:2001at}. If
the chiral expansion would have converged to that order one should see that
this component is actually very much suppressed in the phenomenological
analysis of NN scattering data. The results presented in
Fig.~\ref{fig:mom-sys-local} show that this is not the case. As we see, the
contribution to $W_Q$ (proportional to $V_{14a}$, see
Appendix~\ref{sec:mom-sp}) which vanishes for ChPT to N5LO is not vanishing
within uncertainties. Thus, this is a serious indication that one has still to
go higher orders to accommodate the non-vanishing value character of all NN
potential components. 

Finally, we remind that Eq.~(\ref{eq:on-id}) does not hold off-shell
but to N5LO $V_P=W_P=0$. Such a dependence appears when considering
$A_1$ exchange to ${\cal O}(1/M_N^2)$. Taking into account that the
quantum numbers of $A_1$ are the same as $3\pi$ (The decay process
$A_1 \to \rho \pi \to (\pi\pi) \pi$ is the dominant mechanism ) and
that there are $3\pi$ exchange contributions to N5LO some
clarification would be needed to explain to what order does one expect
a non-vanishing of $V_P$ and $W_P$.

\begin{table*}
\caption{\label{tab:Cut-offdependence}  Fits of the $\chi$-TPE potentials
depending on the cutoff radius and the maximum fitting energy as described
in~\cite{Perez:2013za}. The chiral constants of the fourth line were taken
from a Ref.~\cite{Ekstrom:2013kea,Ekstrom:2014dxa} and used as fixed values
during the $\chi^2$ minimization with respect of the delta-shell parameters.
Highest counter-term column indicates the maximum angular momentum where at
least one delta-shell strength coefficient is non-vanishing.}
% Give a unique label
\centering
\begin{tabular}{l c D{.}{.}{3.6} 
        D{.}{.}{2.5} D{.}{.}{1.4}  c c}
      \hline\noalign{\smallskip}
      Max $T_{\rm LAB}$ & $r_c$ & 
      \multicolumn{1}{c}{$c_1$} & 
      \multicolumn{1}{c}{$c_3$} & 
      \multicolumn{1}{c}{$c_4$} & Highest  & $\chi^2/\nu$ \\
      MeV & fm & 
      \multicolumn{1}{c}{GeV$^{-1}$} & 
      \multicolumn{1}{c}{GeV$^{-1}$} & 
      \multicolumn{1}{c}{GeV$^{-1}$} & counter-term & \\
      \noalign{\smallskip}\hline\noalign{\smallskip}
      350 & 1.8 & -0.4(11) & -4.7(6)  & 4.3(2) & $F$ & 1.08 \\
      350 & 1.2 & -9.8(2)  &  0.3(1)  & 2.84(5)& $F$ & 1.26 \\
      125 & 1.8 & -0.3(29) & -5.8(16) & 4.2(7) & $D$ & 1.03 \\
      125 & 1.2 & -0.92    & -3.89    & 4.31   & $P$ & 1.70 \\
      125 & 1.2 & -14.9(6) & 2.7(2)   & 3.51(9)& $P$ & 1.05 \\
      \noalign{\smallskip}\hline
\end{tabular}
\end{table*}

\section{Chiral Miscellany and loose ends}

\subsection{Naturalness}

While the statistical approach to NN is based on a standard least
squares optimization, which minimizes the distance between the theory
and the experiment for many pp and np scattering data, it is important
to underline that not all fits are eligible and in fact some of them
should be rejected, either because they generate spurious bound states
or because some of the fitting parameters turn out to be rather
different than grounded theoretical expectations.

In the case of $\chi$TPE exchange, the chiral constants $c_{1,3,4}$ are
saturated by meson exchange~\cite{Bernard:1996gq}. Actually, $c_1$ is
saturated by scalar exchange $c_1^S = - g_S c_m /m_S^2 $. Taking $M_N = g_S
F_\pi $ and $c_m = F_\pi/2$ , $m_S=m_V = F_\pi \sqrt{24 \pi /N_c}$ and $M_N =
N_c m_\rho/2$~\cite{Ledwig:2014cla} we get $c_1^S \sim -N_c /( 4 \sqrt{2}
m_\rho) \sim -0.7 {\rm GeV}^{-1} $ and $c_3$ and $c_4$ are saturated by
$\Delta$ resonance; taking $\Delta = M_\Delta -M_N$ one has $c_2^\Delta =
-c_3^\Delta = 2 c_4^\Delta = g_A^2/(2 \Delta) \sim 2.97 {\rm GeV}^{-1}$. Of
course, these are not very accurate values, but indicate the order of
magnitude one should expect.

The results in Table~\ref{tab:fpf0fc-TPE} illustrate this point when the
$\chi$TPE (N2LO) potential is used to fit the Granada-2013 database and the
chiral constants $c_{1,3,4}$ are taken as fitting parameters.  As we see, the
best fit {\it does not} produce natural values for $c_{1,3,4}$ and hence
should be rejected. The natural values correspond to the cut-off distance
$r_c=1.8-2.4 {\rm fm}$. The table also illustrates that the small cut-off
$r_c=1.2 {\rm fm}$ is incompatible with the data. From the point of view of
the effective elementarity of the nucleon, where $r_e=1.8$fm, the value
$r_c=1.8$fm is the {\it smallest possible} and hence provides the minimal
number of parameters, which for pn+pp turns out to be about $N_{\rm Par}=30$
when $T_{\rm LAB} \le 350$MeV.

\subsection{Power counting vs Coarse graining}

We want to elaborate on the main differences between power counting
and coarse graining. As mentioned before, in the partial wave
amplitude we have a left-hand branch cuts due to multiple $n-\pi$
exchanges, $p,p' = i n m_\pi /2 $ Thus, in a low momentum expansion
corresponding to $|p|,|p'| < m_\pi /2 $ we have a Taylor expansion as
in Eq.~(\ref{eq:vlowk}), but the coefficients are functions of
$m_\pi$. If we count powers of $Q=p,p',m_\pi$ we have further the
expansion
\begin{equation}
C_0(m_\pi^2) = c_0 + c_0' m_\pi^2 + \dots 
\end{equation}
Already at this level we see the phenomenon of parameter redundancy;
in a combination such as $c_0 + c_0' m_\pi^2 + \dots $ the constants
$c_0$, $ c_0' $, etc. {\it cannot be } disentangled from NN data. As a
general rule, the number of parameters of the pion-less theory is
smaller than the pion-full theory, since the pion exchange potentials
also contain LEC's (stemming e.g.  from $\pi N$ scattering). Thus,
while a power counting scheme the total number of parameters depends
on the the maximal energy {\it and} the accuracy of the experimental
data, in the coarse graining scheme the number of parameters is
essentially limited by the maximal fitting energy, see
Eq.~(\ref{eq:npar}), fixing the discernible wavelength resolution.

\subsection{The short distance cut-off}
\label{sec:sdc}
The questions on the numerical cut-of $r_c$ supported by the analysis of data
can be resolved by explicitly separating the potential as indicated by
Eq.~(\ref{eq:sep}) and changing the maximal fitting energy. The results checked
to be statistically consistent in ~\cite{Perez:2013za} are summarized in
Table~\ref{tab:Cut-offdependence}. It is striking that $D$-waves, nominally
N3LO and forbidden by Weinberg chiral counting at N2LO, are indispensable !.
Furthermore, data and N2LO do not support $r_c < 1.8{\rm fm}$, while several
$\chi$-potentials~\cite{Gezerlis:2014zia,Piarulli:2014bda} take $r_c=0.9-1.1
{\rm fm}$ as ``reasonable''. This need of higher order counter-terms is also
found in more complete and higher order calculations , where $F$-waves (N5LO)
counter-terms are added despite keeping the pion exchange contribution to one
lower order.

Of course, one may think that 125 MeV is too large an energy. We find that
when we go down to 40 MeV, the $\chi$TPE potential becomes invisible being
compatible with zero within statistical uncertainties from the experimental
data~\cite{Amaro:2013zka,Perez:2013cza}.

The chiral potential (including $\Delta$-degrees of freedom) of
Ref.~\cite{Piarulli:2014bda} explicitly violates Weinberg's counting
since it has N2LO long distance and N3LO short distance pieces, and
residuals are not Gaussian. More recently, the local short distance
components of this potential have been fitted up to 125 MeV LAB
energy~\cite{Piarulli:2016vel} improving the goodness of the fit,
similarly to~\cite{Perez:2013za} (see also table
\ref{tab:Cut-offdependence}). As already said, the most recent
and improved N4LO chiral potentials by Idaho-Salamanca~\cite{Entem:2017gor} and the Bochum~\cite{Reinert:2017usi} groups 
suffer also from this deficiency although the order mismatch is
displaced to higher orders. In both cases, a substantial improvement
is achieved by introducing F-wave counter-terms, which properly
speaking are N5LO.

\subsection{The deconstruction argument}

An alternative way of checking the failure of the power counting is provided
by a deconstruction argument~\cite{Perez:2013za}. This corresponds to
determine under what conditions are the short distance phases $\delta_{\rm
short} $, .i.e. the phase shifts stemming solely from $V_{\rm short}$
compatible with zero within uncertainties, i.e. $|V_{\rm short}| < \Delta V$
?.  This is equivalent to determine what  partial waves fulfill $ | \delta_{\rm
  short} | \le \Delta \delta_{\rm stat}$ when $r_c = 1.8 \, {\rm fm} $.
Unfortunately, this does not work for D-waves, when the fit to 125 MeV is
undertaken, in full agreement with the previous conclusions.

\subsection{The impact parameter argument}

The long distance character of $\chi$TPE makes peripheral phases
(large angular momentum) to be suitable for a perturbative comparison
{\it without}
counter-terms~\cite{Kaiser:1997mw,Kaiser:1998wa,Entem:2014msa}. However,
one should take into account that 1) peripheral phases can only be
obtained from a complete phase shift analyses and 2) their
uncertainties are tiny~\cite{Perez:2013jpa}. The analysis of
\cite{Entem:2014msa} just makes an eyeball comparison which looks
reasonable but the agreement was not quantified\footnote{This was done
  using the SAID database (\url{http://gwdac.phys.gwu.edu/}), a
  $25\sigma$ incompatible fit with $p \ll 1$(see
  e.g.~\cite{Perez:2014waa}).}.  A careful comparison reflects that
the thickness of their points in their figures are {\it much larger}
than the estimated errors from high quality fits. We find that
peripheral waves predicted by 5th-order chiral perturbation theory are
not consistent with the Granada-2013 self-consistent NN database
\cite{RuizSimo:2017anp}. A very vivid way of presenting the
discrepancy is by comparing the phase-shifts in terms of the impact
parameter, defined as
\begin{equation}
  b = (L+ \frac12 ) / p 
\end{equation}
and the variable for every partial wave  
\begin{equation}
\xi^{\rm N4LO}  = \frac{\delta^{\rm N4LO} - {\rm Mean}(\delta)}{{\rm Std}(\delta)}  \, , 
\end{equation}
which measures the deviation from the phase-shift corresponding to the N5LO
phases to the averaged phase-shift divided by its standard deviation. The
conclusions of Ref.~\cite{RuizSimo:2017anp} are that for $ 2 {\rm fm} \le b
\le 5 {\rm fm}$, one has $\xi^{\rm N4LO} \sim 5 $ for the Granada PWA,
$\xi^{\rm N4LO} \sim 3 $ for the 6 Granada potentials, $\xi^{\rm N4LO} \sim 1
$ for the 13 high quality potentials. Our interpretation of this result is
that the {\it systematic} error of the perturbative N5LO reflects the
variations of the peripheral phases in the last 26 years\footnote{Similar
trends are found by the SAID analysis~\cite{Workman:2016ysf} as discussed in
Ref.~\cite{RuizSimo:2017anp}. This possibly due to the rational representation
of the phase-shifts as a function of the energy which has larger discrepancies
than the statistical errors.}.
\begin{equation}
\Delta \delta^{\rm PWA, syst} \sim |\delta^{\chi{\rm,N4LO}}- \delta^{\rm PWA}| \gg  \Delta \delta^{\rm PWA, stat}  \, . 
\end{equation}
Sometimes we get even $3\sigma$ discrepancies.   More details on this
peripheral analysis are presented  in Ref.~\cite{RuizSimo:2017anp}.

\begin{table*}
\caption{\label{tab:LEPS-stat-sys-pert} Low energy threshold np parameters for
all partial waves with $ 2 \leq  j \leq 5$. The central value and
\emph{statistical} error bars are given on the first line using the DS-OPE
potential~\cite{Perez:2013mwa,Perez:2013jpa}.  The second line quotes the
\emph{systematic} uncertainties, the central value and error bars correspond
to the mean and standard deviation of the 6 realistic potentials
NijmII~\cite{Stoks:1994wp}, Reid93~\cite{Stoks:1994wp},
AV18~\cite{Wiringa:1994wb}, DS-OPE~\cite{Perez:2013mwa,Perez:2013jpa},
DS-$\chi$TPE~\cite{Perez:2013oba,Perez:2013cza} and
Gauss-OPE~\cite{Perez:2014yla}. The third line is the N2LO \emph{perturbative}
$\chi$PT result with OPE (charge independent with $f_{\pi NN}^2=0.075 $ and
$m_\pi = (m_{\pi^0}+2m_{\pi^+})/3$) and $\chi$TPE potentials.  For each
partial wave we show the scattering length $\alpha$ and the effective range
$r_0$, both in ${\rm fm}^{l+l'+1}$, as well as the curvature parameters
$v_{2,3,4}$ in ${\rm fm}^{l+l'+3,5,7}$. For the coupled channels we use the
nuclear bar parameterization of the $S$ matrix. Uncertainties smaller than
$10^{-3}$ are not quoted}
{\scriptsize
\begin{tabular*}{\textwidth}{@{\extracolsep{\fill}} l D{.}{.}{1.5} D{.}{.}{1.5}
 D{.}{.}{1.5} D{.}{.}{1.5} D{.}{.}{1.5} D{.}{.}{2.7} }
\hline\noalign{\smallskip}
Wave & {\rm Scheme}  & \multicolumn{1}{c}{$\alpha$} 
        & \multicolumn{1}{c}{$ r_0$} 
        & \multicolumn{1}{c}{$ v_2$} 
        & \multicolumn{1}{c}{$ v_3$} 
        & \multicolumn{1}{c}{$ v_4$} \\
\noalign{\smallskip}\hline\noalign{\smallskip}
% \begin{comment}
% $^1S_0$ &   -23.735(6)  &    2.673(9)  &    -0.50(1)  &     3.87(2)  &   -19.6(1)   \\
%         &   -23.740(6) &     2.69(3)  &    -0.48(1)  &     3.6(1)   &   -19.4(6)   \\
% $^3P_0$ &    -2.531(6)  &    3.71(2)   &     0.93(1)  &     3.99(3)  &    -8.11(5)  \\
%         &    -2.6(2)   &     3.6(5)   &     0.8(6)   &     3.9(1)   &    -8.4(10)  \\
% $^1P_1$ &     2.759(6)  &   -6.54(2)   &    -1.84(5)  &     0.41(2)  &     8.39(9)  \\
%         &     2.77(3)  &    -6.5(1)   &    -1.7(2)   &     0.5(3)   &     8.1(3)   \\
% $^3P_1$ &     1.536(1)  &   -8.50(1)   &     0.02(1)  &    -1.05(2)  &     0.56(1)  \\
%         &     1.53(1)  &    -8.55(8)  &    -0.02(4)  &    -1.0(2)   &     0.3(4)   \\
% $^3S_1$ &     5.435(2)  &    1.852(2)  &    -0.122(3) &     1.429(7) &    -7.60(3)  \\
%         &     5.42(1)  &     1.84(1)  &    -0.132(8) &     1.44(1)  &    -7.65(6)  \\
% $\EP_1$ &     1.630(6)  &    0.400(3)  &    -0.266(5) &     1.47(1)  &    -7.28(2)  \\
%         &     1.62(2)  &     0.40(1)  &    -0.27(1)  &     1.46(2)  &    -7.30(4)  \\
% $^3D_1$ &     6.46(1)   &   -3.540(8)  &    -3.70(2)  &     1.14(2)  &    -2.77(2)  \\
%         &     6.44(4)  &    -3.56(2)  &    -3.76(5)  &     1.09(4)  &    -2.7(1)   \\
% \end{comment}
$^1D_2$ & {\rm PWA}_{\rm stat} &    -1.376     &   15.04(2)   &    16.68(6)  &   -13.5(1)   &    35.4(1)   \\
        & {\rm PWA}_{\rm syst} &    -1.380(7) &    15.0(1)   &    16.6(2)   &   -13.1(3)   &    36.1(16)  \\
        & {\rm N2LO}_{\rm pert} &    -1.768    &    12.81      & 12.92 & -18.36 & -3.602 \\ \hline 
$^3D_2$ & {\rm PWA}_{\rm stat} &    -7.400(4)  &    2.858(3)  &     2.382(9) &    -1.04(2)  &     1.74(2)  \\
        & {\rm PWA}_{\rm syst} &    -7.40(1)  &     2.861(8) &     2.40(2)  &    -0.98(3)  &     1.8(1)   \\
	& {\rm N2LO}_{\rm pert}	& -7.564 & 2.784 & 2.323 & -0.931 & 1.775 \\ \hline 
$^3P_2$ & {\rm PWA}_{\rm stat} &     -0.290(2)  &   -8.19(1)   &    -6.57(5)  &    -5.5(2)   &   -12.2(3)   \\
        & {\rm PWA}_{\rm syst} &    -0.287(6) &    -8.2(2)   &    -6.6(7)   &    -5.3(18)  &   -11.7(24)  \\
	& {\rm N2LO}_{\rm pert}	& 1.028 & -3.346 & 4.157 & -3.359 & -11.56 \\ \hline 
$\epsilon_2$ & {\rm PWA}_{\rm stat} &      1.609(1)  &  -15.68(2)   &   -24.91(8)  &   -21.9(3)   &   -64.1(7)   \\
        & {\rm PWA}_{\rm syst} &     1.607(6) &   -15.7(2)   &   -25.0(7)   &   -21.9(30)  &   -63.6(69)  \\
	& {\rm N2LO}_{\rm pert}	& 2.039 & -6.480 & -0.457 & 13.02 & -43.64 \\ \hline 
$^3F_2$ & {\rm PWA}_{\rm stat} &    -0.971     &   -5.74(2)   &   -23.26(8)  &   -79.5(4)   &  -113.0(16)  \\
        & {\rm PWA}_{\rm syst} &    -0.970(5) &    -5.7(1)   &   -23.1(6)   &   -79.0(35)  &  -113.0(129) \\
	& {\rm N2LO}_{\rm pert}	& -1.371 & 7.416 & 24.53 & 49.59 & -37.49 \\ \hline 
$^1F_3$ & {\rm PWA}_{\rm stat} &      8.378     &   -3.924     &    -9.869(4) &   -15.27(2)  &    -1.95(7)  \\
        & {\rm PWA}_{\rm syst} &     8.376(7) &    -3.927(5) &    -9.89(3)  &   -15.4(2)   &    -2.3(4)   \\
        & {\rm N2LO}_{\rm pert} & 8.531 & -3.836 & -9.620 & -14.75 & -1.51 \\  \hline 
$^3F_3$ & {\rm PWA}_{\rm stat} &      2.689     &   -9.978(3)  &   -20.67(2)  &   -19.12(8)  &   -27.7(2)   \\
        & {\rm PWA}_{\rm syst}&     2.692(7) &    -9.97(3)  &   -20.6(1)   &   -19.0(4)   &   -27.0(7)   \\
        & {\rm N2LO}_{\rm pert} & 3.779 & -8.142 & -18.25 & -21.34 & -0.773 \\  \hline 
$^3D_3$ & {\rm PWA}_{\rm stat}  &    -0.134     &    1.373     &     2.082(3) &     1.96(1)  &    -0.45(3)  \\
        & {\rm PWA}_{\rm syst} &    -0.15(2)  &     1.370(3) &     2.07(2)  &     1.91(7)  &    -0.5(1)   \\
        & {\rm N2LO}_{\rm pert} & -0.228 & 1.330 & 1.986 & 1.768 & -0.857  \\    \hline 
$\epsilon_3$ & {\rm PWA}_{\rm stat} &     -9.682     &    3.262     &     7.681(3) &     9.62(2)  &    -1.09(5)  \\
        & {\rm PWA}_{\rm syst} &    -9.684(6) &     3.258(5) &     7.66(3)  &     9.5(1)   &    -1.2(2)   \\
        & {\rm N2LO}_{\rm pert} & -9.847 & 3.170 & 7.39 & 8.79 & -2.61 \\    \hline 
$^3G_3$ & {\rm PWA}_{\rm stat} &    4.876     &   -0.027     &     0.019(2) &     0.07(1)  &    -2.69(3)  \\
        & {\rm PWA}_{\rm syst} &     4.875(3) &    -0.04(1)  &    -0.03(6)  &    -0.2(3)   &    -3.1(6)   \\
        & {\rm N2LO}_{\rm pert} & 4.941 & -0.084 & -0.342 & -2.18 & -9.89 \\  \hline 
$^1G_4$ & {\rm PWA}_{\rm stat} &     -3.208     &   10.833(1)  &    34.629(9) &    83.04(8)  &   108.1(4)   \\
       & {\rm PWA}_{\rm syst} &    -3.213(8) &    10.81(2)  &    34.54(8)  &    82.5(4)   &   106.1(17)  \\
       & {\rm N2LO}_{\rm pert} & -5.177 & 7.889 & 27.79 & 77.233 & 66.44 \\   \hline 
$^3G_4$ & {\rm PWA}_{\rm stat} &    -19.145     &    2.058     &     6.814    &    16.769(4) &    10.00(2)  \\
       & {\rm PWA}_{\rm syst} &   -19.15(1)  &     2.058(1) &     6.814(4) &    16.78(2)  &    10.05(7)  \\
       & {\rm N2LO}_{\rm pert} & -19.409 & 2.020 & 6.672 & 16.406 & 9.85 \\   \hline 
$^3F_4$ & {\rm PWA}_{\rm stat} &     -0.006     &   -3.043     &    -4.757(1) &    73.903(5) &   662.21(9)  \\
       & {\rm PWA}_{\rm syst} &    -0.009(3) &    -3.040(8) &    -4.75(5)  &    74.0(4)   &   662.5(40)  \\
       & {\rm N2LO}_{\rm pert} & -0.0136 & -2.504 & -7.15 & -14.03 & -4.24 \\  \hline 
$\epsilon_4$ & {\rm PWA}_{\rm stat} &     3.586     &   -9.529     &   -37.02(3)  &  -184.40(2)  &  -587.28(9)  \\
       & {\rm PWA}_{\rm syst} &     3.590(9) &    -9.53(2)  &   -37.02(7)  &  -184.5(3)   &  -586.4(19)  \\
       & {\rm N2LO}_{\rm pert} & 5.783 & -6.954 & -23.92 & -64.11 & -53.56 \\  \hline 
$^3H_4$ & {\rm PWA}_{\rm stat} &    -1.240     &   -0.157(2)  &    -1.42(1)  &   -14.0(1)   &   -99.0(9)   \\
       & {\rm PWA}_{\rm syst} &    -1.241(4) &    -0.17(1)  &    -1.51(9)  &   -14.9(9)   &  -105.4(59)  \\
       & {\rm N2LO}_{\rm pert} & -2.351 & -0.108 & -1.088 & -8.58 & -30.63 \\  \hline 
$^1H_5$ &  {\rm PWA}_{\rm stat} &   28.574     &   -1.727     &    -7.906    &   -32.787    &   -59.361    \\
       & {\rm PWA}_{\rm syst} &    28.58(2)  &    -1.727    &    -7.905(4) &   -32.78(2)  &   -59.38(6)  \\
	& {\rm N2LO}_{\rm pert}	& 28.830 & -1.701 & -7.758 & -32.09 & -58.03 \\ \hline 
$^3H_5$ &   {\rm PWA}_{\rm stat} &   6.081     &   -6.439     &   -25.228    &   -82.511(3) &  -168.47(2)  \\
       & {\rm PWA}_{\rm syst} &     6.09(2)  &    -6.43(2)  &   -25.21(6)  &   -82.5(1)   &  -168.1(9)   \\
	& {\rm N2LO}_{\rm pert}	& 11.53 & -4.135& -18.10 & -70.12 & -112.1 \\  \hline 
$^3G_5$ &   {\rm PWA}_{\rm stat} &  -0.008     &    0.481     &     1.878    &     6.100    &     6.791    \\
       & {\rm PWA}_{\rm syst} &    -0.010(2) &     0.480    &     1.878(1) &     6.098(4) &     6.78(1)  \\
	& {\rm N2LO}_{\rm pert}	& -0.009 & 0.471 & 1.837 & 5.958 & 6.63 \\  \hline 
$\epsilon_5$ & {\rm PWA}_{\rm stat} &   -31.302     &    1.556     &     6.995    &    28.179    &    48.376(2) \\
       & {\rm PWA}_{\rm syst} &   -31.31(2)  &     1.556    &     6.993(4) &    28.17(1)  &    48.35(3)  \\
	& {\rm N2LO}_{\rm pert}	& -31.581 & 1.533 & 6.863 & 27.58 & 47.29 \\  \hline 
$^3I_5$ &     {\rm PWA}_{\rm stat} & 10.678     &    0.011     &     0.146    &     1.441    &     6.546(6) \\
       & {\rm PWA}_{\rm syst} &    10.680(6) &     0.011    &     0.144(1) &     1.43(2)  &     6.5(1)   \\
	& {\rm N2LO}_{\rm pert}	& 10.711 & 0.010 & 0.143& 1.41 & 6.35 \\
\noalign{\smallskip}\hline
\end{tabular*}
}
\end{table*}

\subsection{Threshold parameters at N2LO: Perturbative vs Non-perturbative}

If chiral perturbation theory works for NN, the best scenario to check it is,
as already mentioned, by looking at long distance properties, namely large
angular momenta and small energies. This comparison provides a stringent test
on the quality of chiral potentials. To this end we use the effective range
expansion of the inverse amplitude generalized to higher partial waves
including coupled channels, see, Eq.~(\ref{eq:ERE-coup}), going to ${\cal O}
(p^8)$. Practical formulas for the low energy threshold parameters have been
deduced some time ago for the NijmII and Reid93
potentials~\cite{PavonValderrama:2005ku} and extended more recently for the
set of 6 Granada potentials~\cite{RuizSimo:2017anp} using a discretized form
of the variable phase approach of Calogero~\cite{calogero_Variable_1967}.

We can illustrate the perturbative character of peripheral waves by
comparing the low energy parameters, computed in chiral perturbation
theory by using the N2LO TPE as a reference. This has the additional
advantage that due to the angular momentum suppression there is a
large insensitivity to the short distances so that the limit $r_c \to
0$ can be safely taken. However, it was shown in
Ref.~\cite{PavonValderrama:2005uj} that due to the short distance
singularities of the chiral potentials, see Eq.~(\ref{eq:chi-short}),
there is a finite order in perturbation theory where regardless of the
angular momentum the result is divergent. Possibly the most direct way
to approach the calculation in perturbation theory is via the variable
phase approach.  Analytical formulas can be obtained but they are too
lengthy to be quoted here, so we just give the final outcome and
comment it. For the N2LO TPE potential the first perturbative
calculable partial waves are D-waves, and for increasing F-,G-,H-,
etc.  partial waves we expect a faster convergence. In
Table~\ref{tab:LEPS-stat-sys-pert} we show the results when we take
the central values of the chiral constants found in our previous
work~\cite{Perez:2013oba}. As we generally see from a dedicated
inspection of the table, while the perturbative numbers get closer to
the full result for increasing angular momentum, including the
systematic spread of the 6 Granada potentials, there is still a
significant discrepancy due to the finite order of the
perturbation. We remind that the same N2LO TPE used above $r_c =1.8$fm
{\it to all orders} together with a delta-shells potential yields a
satisfactory description of full Granada-2013 database.

\subsection{Counter-terms vs Renormalization conditions: The zero energy argument}

The low energy threshold parameters allow to probe the structure of chiral
potentials against the NN interaction. The current approach to chiral
interactions is to incorporate the $\chi$TPE tail and include short range
counter-terms fitted to pp and np phase-shifts or scattering
data~\cite{Ekstrom:2013kea,Ekstrom:2014dxa}\footnote{In momentum space
counter-terms corresponds to coefficients of polynomials, see e.g. 
\cite{RuizArriola:2016vap}, which can be fixed by low energy threshold
parameters by implicit renormalization.}.  However, these approaches are
subject to strong systematic uncertainties since a fit to phase-shifts may be
subjected to off-shell ambiguities and so far low energy chiral potentials
fitted to data have not achieved Gaussian residuals~\cite{Ekstrom:2014dxa} or
even have huge~\cite{Gezerlis:2014zia} to moderate \cite{Piarulli:2014bda}
$\chi^2/\nu$ values. To avoid these shortcomings we use
$\chi$TPE~\cite{Perez:2013oba,Perez:2013cza} with a simpler short range
structure inferred from low energy threshold parameters~\cite{Perez:2014waa}
with their uncertainties inherited from the 2013-Granada
fit~\cite{Perez:2013jpa}. This corresponds to zero energy renormalization
condition of the counter-terms.

One could naively expect to be able to set any number of short range
counter-terms to reproduce the same number of low energy threshold parameters.
Actually, in order to have the 9 counter-terms dictated by Weinberg to N2LO as
in~\cite{Ekstrom:2013kea} we need to fix $\alpha_0$ and $r_0$ for both $^1S_0$
and $^3S_1$ waves, the mixing $\alpha_\epsilon$ and $\alpha_1$ for the
$^3P_0$, $^3P_1$, $^3P_2$,$^1P_1$~\cite{Perez:2014waa}. In practice this
turned out to be unfeasible in particular for the $J=1$ coupled channel where
one has matrices $\mathbf{a}$ and $\mathbf{r}_0$. If instead one includes two
counter-terms in each partial wave in the $J=1$ coupled channel it is then
possible to reproduce the coupled channel $\mathbf{a}$ and $\mathbf{r}_0$
matrices.  With this structure we have a total of 12 short range parameters
set to reproduce 12 low energy threshold parameters from~\cite{Perez:2014waa},
and not the 9 expected from N2LO~\cite{Ekstrom:2013kea}. Nonetheless a good
description of the phase-shifts up to a laboratory energy of $20$ MeV is
observed~\cite{RuizArriola:2016sbf}.

\section{Outlook: To count or not to count}

%\begin{comment}

%\end{comment}

Chiral nuclear forces have emerged in Nuclear Physics providing a
unified description of nuclear phenomena more rooted in QCD and less
model dependent than most of the phenomenological approaches and with
a chiral hierarchy in the different effects, including multi-nucleon
forces, as suggested initially by Weinberg. The huge computational
effort of the chiral nuclear agenda proves that they are not only
calculable but also that they can be used in light and medium-mass
nuclei studies. Despite the theoretical appeal and promising
developments in the last 30 years it is fair to say that their
indispensability remains to be established and we believe that further
efforts must be taken in this direction in order to possibly
consolidate their significance. With this motivation in mind we have
analyzed some issues which go into the validation vs falsification of
the latest most accurate N4LO chiral potential fits. In fact, a
significant step forward in the NN case has been made by the Bochum
group which has shown that several non-chiral high quality NN
potentials such as AV18 and NijII are statistically less quality by
chiral potentials with less parameters which are in between N4LO and
N5LO (N4LO$^+$) when a fit to the Granada 2013 (with some additional
data exclusion and lower energies) is carried out.  A similar
situation emerges from the Idaho-Salamanca analysis which also works
at the in-between N4LO$^+$ scheme (plus data re-shuffling), but does
not outperform the non-chiral potentials.  One worrisome aspect is
that the large difference may be due to systematic uncertainties.
Within the EFT approach there is a residual model dependence regarding
the finite cut-off regularization scheme. In our experience with
6-Granada potentials fitting the same database in a statistically
satisfactory manner, the parametrization of the short range component
of the nuclear force, $r \le 1.8$fm dominates the uncertainties.  We
stress that none of these results invalidates using $\chi$TPE, say at
N2LO above $r_c=1.8 {\rm fm}$ and describing about 7000 NN data with
30 parameters, but it does question the status of Weinberg's power
counting encoding the short distance component of the interaction when
facing NN data. This analysis, however, does not shed any obvious
light on the lasting discussions about its consistency.

The best possible validation of chiral nuclear forces would occur if
they could be used {\it themselves} as a reliable tool to {\it fit and
  select} scattering data. The fulfillment of such a goal would be a
major achievement for the theory. We report indications that for the
present there still some, hopefully small, way to go to come to a such
a situation. Our statistical, perturbative and peripheral analyses do
not indicate otherwise.  For instance, one quadratic spin component of
the NN potential, $W_Q$, vanishes to N5LO while all high quality
potentials provide a small but significantly non-vanishing component
within uncertainties. If we assume that the next N6LO order will do
the job the number of parameters then becomes comparable to that of
the non-chiral potentials.  Thus, one needs to check if chiral forces
might not be necessarily {\it more predictive} than the usual
phenomenological and non-chiral approaches.

\begin{table}[ttt]
\caption{\label{tab:op-mom} Argonne V18  momentum-space spin-isospin operators}
{\small \begin{tabular*}{\columnwidth}{@{\extracolsep{\fill}}ll}
\hline\noalign{\smallskip}
Term & spin-isospin operator \\
\noalign{\smallskip}\hline\noalign{\smallskip}
$\tilde{O}_{1}$ & $ \mathbf{I} $ \\
$\tilde{O}_{2}$ & $ (\btau_1 \cdot \btau_2) $\\
$\tilde{O}_{3}$ & $ (\pmb{\sigma}_1 \cdot \pmb{\sigma}_2)$\\
$\tilde{O}_{4}$ & $(\pmb{\sigma}_1 \cdot \pmb{\sigma}_2)(\btau_1 \cdot \btau_2)$\\ 
$\tilde{O}_{5}$ & $ - \left (3 (\mathbf{q} \cdot \pmb{\sigma}_1)( \mathbf{q} \cdot \pmb{\sigma}_2)
- {q^2 } \pmb{\sigma}_1 \cdot \pmb{\sigma}_2 \right ) $\\
$
\tilde{O}_{6} $ & $
- \left (3 (\mathbf{q} \cdot \pmb{\sigma}_1)( \mathbf{q} \cdot \pmb{\sigma}_2)
- {q^2 } \pmb{\sigma}_1 \cdot \pmb{\sigma}_2 \right )
(\btau_1 \cdot \btau_2) $\\
$
\tilde{O}_{7} $ & $
i (\mathbf{k} \times \mathbf{k}')\cdot \mathbf{S} $\\  
$ \tilde{O}_{8} $ & $
i (\mathbf{k} \times \mathbf{k}') \cdot \mathbf{S} (\btau_1 \cdot \btau_2)
$\\
$
\tilde{O}_{9a} $ & $
-(\mathbf{k}'\times \mathbf{k}) \cdot 
(\mathbf{k}'\times \mathbf{k})  
$ \\
$
\tilde{O}_{9b} $ & $
2 (\mathbf{k}'\cdot \mathbf{k}) $\\ 
$
\tilde{O}_{10a} $ & $
-(\mathbf{k}'\times \mathbf{k}) \cdot 
(\mathbf{k}'\times \mathbf{k})  (\btau_1 \cdot \btau_2)
$\\
$
\tilde{O}_{10b} $ & $
2 (\mathbf{k}'\cdot \mathbf{k})(\btau_1 \cdot \btau_2) 
$\\
$
\tilde{O}_{11a}$ & $
-(\mathbf{k}'\times \mathbf{k}) \cdot 
(\mathbf{k}'\times \mathbf{k}) (\pmb{\sigma}_1 \cdot \pmb{\sigma}_2) 
$\\
$
\tilde{O}_{11b} $ & $
2(\mathbf{k}'\cdot \mathbf{k})(\pmb{\sigma}_1 \cdot \pmb{\sigma}_2)  
$\\
$
\tilde{O}_{12a} $ & $
-(\mathbf{k}'\times \mathbf{k}) \cdot 
(\mathbf{k}'\times \mathbf{k})  (\pmb{\sigma}_1 \cdot
\pmb{\sigma}_2) (\btau_1 \cdot \btau_2) 
$\\
$
\tilde{O}_{12b} $ & $
2 (\mathbf{k}'\cdot \mathbf{k})(\pmb{\sigma}_1 \cdot
\pmb{\sigma}_2) (\btau_1 \cdot \btau_2) 
$\\
$
\tilde{O}_{13a} $ & $
- (\mathbf{S} \cdot ( \mathbf{k}\times \mathbf{k}'))^2
$\\
$\tilde{O}_{13b} $ & $
( \mathbf{k}'\times \mathbf{S})\cdot
( \mathbf{k}\times \mathbf{S}) 
$\\
$
\tilde{O}_{14a} $ & $
- (\mathbf{S} \cdot ( \mathbf{k}\times \mathbf{k}'))^2
 (\btau_1 \cdot \btau_2) 
$\\
$
\tilde{O}_{14b} $ & $
( \mathbf{k}'\times \mathbf{S})\cdot
( \mathbf{k}\times \mathbf{S}) (\btau_1 \cdot \btau_2)
$\\
$\tilde{O}_{15}  $ & $
T_{12} 
$\\
$
\tilde{O}_{16} $ & $
(\pmb{\sigma}_1 \cdot \pmb{\sigma}_2) T_{12} 
$ \\
$
\tilde{O}_{17} $ & $
- \left (3 (\mathbf{q} \cdot \pmb{\sigma}_1)( \mathbf{q} \cdot \pmb{\sigma}_2)
- {q^2 } \pmb{\sigma}_1 \cdot \pmb{\sigma}_2 \right )T_{12}
$\\
$
\tilde{O}_{18} $ & $
(\tau_{1z} + \tau_{2z}). 
$\\
\noalign{\smallskip}\hline
  \end{tabular*}
  }
\end{table}

%\vskip1cm 
\begin{acknowledgement}

We thank M. Pav\'on Valderrama for reading the ms. and J.E. Amaro and
I. Ruiz Sim\'o for discussions. The work of E.R.A.  supported partly
supported by the Spanish Ministerio de Economía y Competitividad and
European FEDER funds (grant FIS2017-85053-C2-1-P) and Junta de
Andaluc\'{\i}a grant FQM-225

\end{acknowledgement}

%\begin{comment}
\appendix

\section{Momentum space components}
\label{sec:mom-sp}

\begin{figure*}[htb]
\begin{center}
\includegraphics[width=0.95\linewidth]{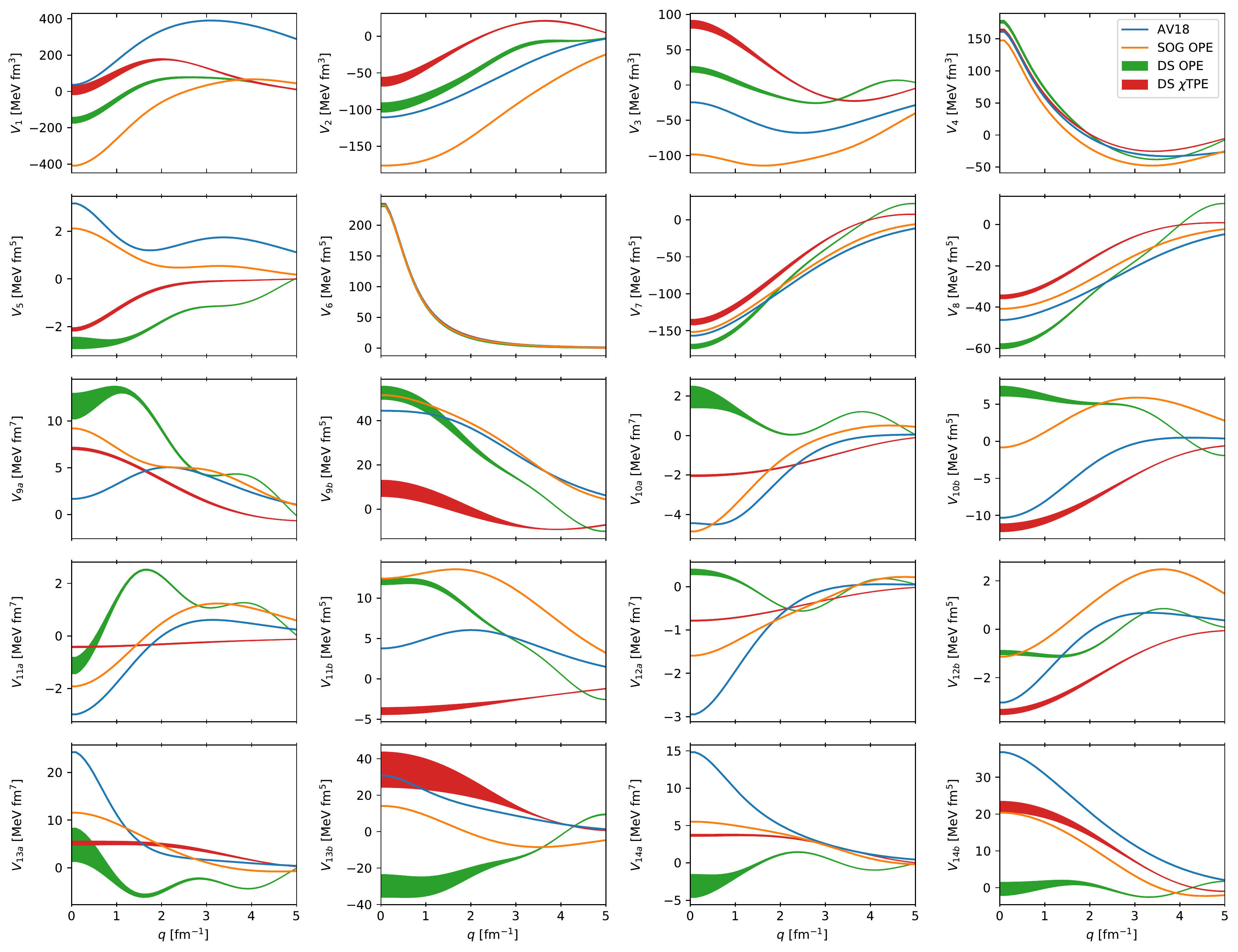}
\end{center}
\caption{\label{fig:mom-sys} Momentum space local components of
  potentials as a function of the momentum transfer for several
  potentials: AV18~\cite{Wiringa:1994wb} and the Granada
  DS-OPE~\cite{Perez:2013mwa,Perez:2013jpa},
  DS-$\chi$TPE~\cite{Perez:2013oba,Perez:2013cza} and
  Gauss-OPE~\cite{Perez:2014yla}.}
\end{figure*}

We use the notation of Ref.~\cite{Veerasamy:2011ak} and quote for
completeness the main results. The basic
Fourier transforms are
\begin{strip}
  \begin{align}
 \int e^{-i \mathbf{k}'\cdot \mathbf{r}}
V (r) e^{i \mathbf{k}\cdot \mathbf{r}} d\mathbf{r} &= 
\int_0^\infty  4 \pi  j_0 (qr)  V (r) r^2 dr . \\
\int e^{-i \mathbf{k}'\cdot \mathbf{r}}
V (r) \mathbf{L} \cdot \mathbf{S}
e^{i \mathbf{k}\cdot \mathbf{r}} d\mathbf{r} &=
i  (\mathbf{k} \times \mathbf{k}')\cdot \mathbf{S} 
\int_0^\infty  \frac{4 \pi }{ q} j_1 (qr) V (r) r^3 dr . \\
\int e^{-i \mathbf{k}'\cdot \mathbf{r}}
V (r) \mathbf{L} \cdot \mathbf{L}
e^{i \mathbf{k}\cdot \mathbf{r}} d\mathbf{r} &=
2 (\mathbf{k}'\cdot \mathbf{k}) \frac
{4\pi }{  q}\int_0^\infty j_1 (qr) V (r) r^3 dr
- (\mathbf{k}'\times \mathbf{k}) \cdot 
(\mathbf{k}'\times \mathbf{k})  
\frac{4 \pi }{ q^2}\int_0^\infty j_2 (qr) V (r) r^4 dr
. \\
\int e^{-i \mathbf{k}'\cdot \mathbf{r}}
V (r) (\mathbf{L} \cdot \mathbf{S})^2
e^{i \mathbf{k}\cdot \mathbf{r}} d\mathbf{r} &= ( \mathbf{k}'\times \mathbf{S})\cdot
( \mathbf{k}\times \mathbf{S})  \frac
{4\pi }{  q} \int_0^\infty j_1 (qr) V_j(r) r^3 dr 
- (\mathbf{S} \cdot ( \mathbf{k}\times \mathbf{k}'))^2 
\frac {4\pi }{  q^2 } \int_0^\infty j_2 (qr) V (r) r^4 dr . \\ 
\int e^{-i \mathbf{k}'\cdot \mathbf{r}}
V(r) S_{12} (\hat r)
e^{i \mathbf{k}\cdot \mathbf{r}} d\mathbf{r} &=
- S_{12}(\hat q) \int_0^\infty 4 \pi j_2 (qr) V(r) r^2 dr . 
  \end{align}
\end{strip}
where $S_{12} (\hat r) = 3(\hat{\mathbf{r}} \cdot \pmb{\sigma}_1)
(\hat{\mathbf{r}} \cdot \pmb{\sigma}_2) - \pmb{\sigma}_1\cdot
\pmb{\sigma}_2$.  In table~\ref{tab:op-mom} $T_{12}$ is the isotensor
operator $T_{12}:= 3 \tau_{1z}\tau_{2z}- \pmb{\tau}_1 \cdot
\pmb{\tau}_2 $.  While the isospin operators, $\pmb{\tau}_i$, factor
out of the Fourier transforms, the operators $L^2$, $\mathbf{L}\cdot
\mathbf{S}$, $(\mathbf{L}\cdot \mathbf{S})^2$ and the tensor operator
$S_{12}$ contribute to the Fourier transform. The potential components
are plotted in Fig.~\ref{fig:mom-sys} and  the relation to Eq.~(\ref{eq_nnamp})
becomes 
\begin{align}
V_C &=  V_1-( V_{9 a} + V_{10 a} + V_{11 a} +V_{12 a} + \frac{1}{2} V_{13 a}) \nonumber \\ 
        & \quad \times\left(P^2 q^2-(q.P)^2\right) + (2 V_{9 b} + V_{13 b}) \left(P^2-\frac{q^2}{4}\right) \\
W_C &=  V_2-( V_{12 a} + \frac{1}{2} V_{14 a})\left(P^2 q^2-(q.P)^2\right) \nonumber \\ 
    &+ (2 V_{10 b} + V_{14 b}) \left(P^2-\frac{q^2}{4}\right) \\
V_S &= V_3 + (2 V_{11 b}+ V_{13 b})P^2 - (\frac{1}{2} V_{11 b}+\frac{1}{4}  V_{13 b}- V_5) q^2 \\
W_S &= V_4 + 2 V_{12 b} \left(P^2-\frac{q^2}{4}\right) + V_{14 b} \left(P^2-\frac{q^2}{4}\right)+q^2 V_6 \\
V_T &= \frac18 V_{13 b}-3 V_5 \\
W_T &= \frac18 V_{14 b}-3 V_6 \\
V_{LS} &= V_7 \\ 
W_{LS} &= V_8 \\ 
V_Q &=-\frac12 V_{13 a}
\end{align}
\begin{align}
W_Q &=-\frac12 V_{14 a} \\
V_P &=- \frac12 V_{13 b} \\
W_P &= -\frac12 V_{14 b}
\end{align}

%\end{comment}

\balance
% BibTeX users please use
%\bibliographystyle{epj}
%\bibliography{biblio}
% \begin{thebibliography}{}
% %
% % and use \bibitem to create references.
% % Format for Journal Reference
% Author, Journal \textbf{Volume}, (year) page numbers.
% Author, \textit{Book title} (Publisher, place year) page numbers
% \end{thebibliography}

\end{document}